\documentclass[11pt,a4paper]{article}
\pdfoutput=1
\usepackage{ifpdf}
\usepackage{jheppub}
\usepackage{rotating}
\usepackage[bb=boondox]{mathalfa}
\usepackage{comment}

\usepackage{tikz-cd}

\usepackage{graphicx} 
\usepackage{mathrsfs}
\usepackage{amsmath,amsfonts,amssymb}
\usepackage{mathtools}
\usepackage{graphicx}
\usepackage{dcolumn}
\usepackage{bm}
\usepackage{hyperref}
\usepackage[mathlines]{lineno}
\usepackage{xcolor}
\usepackage{float}
\usepackage{blindtext}
\usepackage{titlesec}
\usepackage[toc,page]{appendix}
\usepackage{graphicx}
\graphicspath{ {./images/} }
\usepackage[thinc]{esdiff}
\usepackage[euler]{textgreek}
\usepackage{braket}
\usepackage{physics}
\usepackage{xfrac}
\usepackage{soul}
\usepackage{stmaryrd}
\usepackage{pifont}
\usepackage{orcidlink}
\usepackage{enumitem}

\newcommand{\n}{\nonumber}
\renewcommand{\arraystretch}{1.75}
\newcommand{\scO}{\mathcal{O}}
\newcommand{\pd}{\partial}
\newcommand{\Dfbd}{\overset\leftrightarrow{D_{\mu}}}
\newcommand{\Dfb}{\overset\leftrightarrow{D^{\mu}}}

\allowdisplaybreaks

\title{\textcolor{black}{Two-loop dimension Six Effective Action: Integrating Out Heavy Scalar}}

\author[a]{Nilabhra Adhikary,}
\author[a,b]{Jaydeb Das\,\orcidlink{0000-0001-6335-9377},}
\author[a]{Debmalya Dey\,\orcidlink{0009-0004-7475-9850}}

\affiliation[a]{Indian Institute of Technology Kanpur, Kalyanpur, Kanpur 208016, Uttar Pradesh, India.}
\affiliation[b]{Department of Physics, Indian Institute of Technology Guwahati,
North Guwahati, Assam-781039, India.}
\emailAdd{nilabhra09@gmail.com}
\emailAdd{jaydebphys@rnd.iitg.ac.in}
\emailAdd{debmalyad23@iitk.ac.in}

\abstract{For the first time, we present the model-independent two-loop effective action up to dimension six after integrating out heavy scalar(s) employing the Heat-Kernel method. We compute the effective operators that arise at the two-loop level from diagrams involving heavy particles for two example models: heavy electroweak complex triplet and doublet scalars. We present our results on the SILH basis. We also capture the effect in the fermion sector. For these two scenarios, we compute all the fermionic effective operators up to dimension six.}

\begin{document}
\maketitle
\flushbottom

\preprint{}
\newpage
\section{Introduction}\label{sec:intro}
In particle physics, the precision era began with the discovery of the Higgs boson, and experimental measurements have since achieved previously unheard-of levels of accuracy.
Despite the Standard Model (SM) being highly successful in describing fundamental interactions, it remains incomplete, leaving unresolved questions such as the matter-antimatter asymmetry of the universe, the origin of neutrino masses, the hierarchy problem, the strong CP problem, and the nature of dark matter. Beyond Standard Model (BSM) theories offer potential solutions, but the vast landscape of possible models complicates systematic exploration. Effective Field Theory (EFT)\cite{Weinberg:1980wa,Georgi:1994qn,Manohar:2018aog,Cohen:2019wxr}, particularly the Standard Model Effective Field Theory (SMEFT)\cite{Brivio:2017vri,Isidori:2023pyp}, has emerged as a powerful framework to address some of these challenges. By integrating out heavy degrees of freedom and capturing their effects through higher-dimensional operators, EFT enables a model-independent description of new physics while maintaining a direct connection to low-energy observables.
We use SMEFT as a bridge to connect new physics models with experimental data. The framework provides a systematic approach to encode the effects of high-energy phenomena in terms of higher-dimensional operators, which allow for precise low-energy predictions. This approach enhances the understanding of potential BSM signals and offers a way to include quantum corrections, improving the accuracy of predictions for precision observables. This can be done in terms of two perturbative parameters. The first is the energy scale parameter ($\Lambda$), which defines the validity of the EFT below a certain energy threshold. The second is the fine structure constant, which characterizes the loop order, or the level of quantum corrections. As experimental measurements continue to push the boundaries of precision, the need for accurate theoretical predictions becomes more urgent.
Considerable efforts have been dedicated to constructing higher-dimensional effective action operators to account for deviations in low-energy observables and to increase precision regarding the former parameter. Several automated tools have been developed to help in determining the Wilson coefficients of these operators~\cite{Henning:2015alf,Lehman:2015coa,Henning:2017fpj,Lehman:2015via,Fonseca:2019yya,Fonseca:2017lem,Gripaios:2018zrz,Criado:2019ugp,Marinissen:2020jmb,Banerjee:2020bym,Harlander:2023psl,Li:2021tsq,Li:2022tec,BUCHMULLER1986621,Grzadkowski:2010es,Lehman:2014jma,Murphy:2020rsh,Li:2020gnx,Li:2020xlh,Liao:2020jmn,Anisha:2019nzx,Banerjee:2020jun,Hartland:2019bjb,Brivio:2019ius,Brivio:2017btx,Ellis:2020unq,Bagnaschi:2022whn,Ellis:2018gqa,DasBakshi:2021xbl,Naskar:2022rpg,Cepedello:2022pyx,Guedes:2023azv,Gargalionis:2020xvt,Li:2022abx,Chakrabortty:2020mbc}. The inclusion of operators beyond dimension six is essential for capturing subtle UV (ultra-violet) signatures, especially when lower-dimensional operators are absent at tree level. This extended EFT framework helps refine the parameter space of BSM models and ensures that new physics can be effectively explored through precise low-energy observables~\cite{Drozd:2015rsp,Ellis:2016enq,delAguila:2016zcb,Ellis:2017jns,Kramer:2019fwz,Angelescu:2020yzf,Ellis:2020ivx,Bakshi:2018ics,Fuentes-Martin:2022jrf,Carmona:2021xtq,Cohen:2020qvb,Fuentes-Martin:2020udw,Criado:2017khh,Gaillard:1985uh,Cheyette:1987qz,Chan1985,Henning:2014wua,Henning:2016lyp,Dittmaier:2021fls,Fuentes-Martin:2016uol,Cohen:2020fcu,Vandeven:1985, Zhang:2016pja,Banerjee:2023iiv,Chakrabortty:2023yke,Banerjee:2023xak}.
In earlier works~\cite{Banerjee:2023iiv,Chakrabortty:2023yke,Banerjee:2023xak}, the one-loop effective action was computed up to dimension-eight operators for scalar and fermionic theories, including the contributions from heavy-light mixed loops using the HK method~\cite{Minakshisundaram:1949xg,Minakshisundaram:1953xh,Hadamard2003,DeWitt:1964mxt,Seeley:1969,Schwinger:1951nm,Vassilevich:2003xt,Avramidi:2001ns,Avramidi:2015ch5,Kirsten:2001wz,Fulling:1989nb,vonGersdorff:2022kwj,vonGersdorff:2023lle,Carneiro:2024slt}. Building upon this approach, a generalized formalism was proposed in Ref.~\cite{Banerjee:2024rbc} in the context of renormalization of the theory to derive the beta functions for effective operators in scalar and fermionic theories for arbitrary loop orders.
In the Refs.~\cite{Jack:1982hf,Banerjee:2024rbc} the divergent parts of the two-loop calculations were done in detail, and in \cite{Alonso:2022ffe}, a field-space geometric approach was introduced to calculate the same. In this paper, we extend the scope by calculating the two-loop effective action for a generic scalar quantum field theory, considering operators up to dimension six. Our methodology employs the HK spectral approach and builds directly on the framework established in previous studies.
To demonstrate the applicability of this approach, we explore two specific examples: the extension of the SM by an extra Higgs doublet (2HDM) with hypercharge $-1/2$, and the inclusion of an electroweak complex triplet scalar with hypercharge $1$. These two models have distinct phenomenological implications. The 2HDM model~\cite{Anisha:2019nzx,Crivellin:2016ihg,Karmakar:2019vnq,Liu:2015oaa,Dawson:2023ebe,Dawson:2022cmu,Branco:2011iw,Ginzburg:2004vp,Gunion:2005ja} provides sufficient CP-violation (type III), which is an essential ingredient for explaining the baryon asymmetry of the Universe (BAU) via a strong first-order electroweak phase transition (SFOEWPT) and can also generate primordial gravitational wave (GW) signals, all while remaining consistent with electroweak precision data. The complex triplet model, famously known as the type-II Seesaw model~\cite{arhrib:hal-00608687,Schechter:1981cv,Schechter:1980gr,FileviezPerez:2008jbu,Das:2023zby}, explains the generation of neutrino (Majorana) mass, lepton flavor violation processes, and the BAU. Our objective is to develop a framework for calculations that involve the heavy scalar(s) fields that need to be integrated out.  We demonstrate its applicability through examples of two cases with heavy scalar particles in different representations.
Before detailing the structure of our paper, we first highlight the key points of our work, which are as follows. 
\begin{itemize}[topsep=-1ex,itemsep=-0.35ex,partopsep=0.5ex,parsep=0.5ex]
    \item We have computed the two-loop effective action in a model-independent way after integrating out a heavy scalar. Here, we have only considered the loops containing the heavy particle.
    \item We have implemented our results for two cases, which are:
    \begin{enumerate}[topsep=-1ex,itemsep=-0.35ex,partopsep=1ex,parsep=1ex]
           \item SM + electroweak scalar triplet ($\Delta$) with hypercharge $Y_\Delta=1$,
           \item SM + electroweak scalar doublet ($\Phi$) with hypercharge $Y_\Phi=-1/2$.
      \end{enumerate}
\end{itemize}
\paragraph{}
The paper is organized as follows. We briefly overview the HK method for model-independent two-loop effective action calculation in Sec.\,\ref{HK&Green}. We calculate the component Green’s functions (CGFs) and associated algebraic singularities from the interacting Green's function. Then, we demonstrate how to compute different vacuum diagrams with only the heavy particles to capture the two-loop quantum corrections, using the n-point vertex factors obtained from the Lagrangian. We systematically compute the finite part of the vacuum diagrams to get the effective action for a generic Lagrangian involving scalars in Sec.\,\ref{generalU}. We also show that the IR (infra-red) poles (If the massless limit is taken) get canceled after adding all the possible vacuum diagrams with the counter-term diagram, which acts as a sanity check of our calculation.
Next, in Sec.\,\ref{models}, we compute the two-loop corrections for the models, as mentioned earlier. Note that in this work, we restrict ourselves only up to dimension six standard model effective operators.
Finally, we briefly conclude our work in Sec.\,\ref{con}.

There are also contributions from loops containing both the heavy and light degrees of freedom that we have left out of our work. Those heavy-light mixing~\cite{Banerjee:2023xak} contribution for the two-loop will be addressed in our future work.

\section{The interacting scalar Green’s function: Heat-Kernel approach} \label{HK&Green}
Here, we are considering a theory described by a Lagrangian of an $O(n)$ symmetric scalar multiplet $\phi=(\phi_1,\phi_2,...,\phi_n)$ in the presence of some background gauge field $A_\mu$ in space-time dimension \(d\) as follows.
\begin{eqnarray} \label{eq:genlag}
\mathcal{L} &=& \frac{1}{2}\phi^T D^2 \phi + \frac{1}{2} M^2 \phi^2 + V(\phi^2),\text{\hspace{0.3cm}}\text{and} \n \\
\widetilde{\Delta}_{ij} &\equiv& \frac{\delta^2 \mathcal{L}}{\delta \phi_i \delta \phi_j} = (D^2 + M^2) \delta_{ij} + U_{ij}(\phi), \text{\hspace{0.3cm}} \forall \text{\hspace{0.3cm}} i,j=\{1,\cdots, n\}
\end{eqnarray}
where, $M$ is the mass of the scalar field $\phi$, $\phi^2=\phi^T\phi=\sum_{i=1}^n \phi^2_i$ and $D_{\mu} = \partial_{\mu} - i A_{\mu}(x)$ is the covariant derivative. Here, $\widetilde{\Delta}$
\footnote{From now on, we are suppressing the indices for simplicity.}
is a self-adjoint second-order elliptic operator having positive eigenvalues. The operator $U(\phi)$ contains all the information of the potential $V(\phi)$. The Heat Kernel (HK) is defined as the fundamental solution of the heat equation corresponding to the second-order elliptic  operator, which is~\cite{Kirsten:2001wz, Vassilevich:2003xt, Avramidi:2015ch5}
\begin{equation}\label{eq:heat-kernel}
   \mathcal{K} (t,x,y,\widetilde{\Delta}) = \bra{y} e^{-\widetilde{\Delta} t} \ket{x} = \sum_n e^{-\widetilde{\Delta} t}\,\tilde{\phi}_n(x)\,\tilde{\phi}^{\dagger}_n(y),
\end{equation}
where $t$ is a parameter\footnote{Here, $t$ is not to be confused with time.}, and $t>0$ for all possible spacetime points. The functions \( \tilde{\phi}_n \) represent the eigenstates of the elliptic operator \( \Delta \).  As demonstrated in \cite{Osipov:2021dhc, Osipov:2001bj, Banerjee:2023xak}, it is easier to define the HK in the Fourier space. By performing the momentum integral, the HK can be represented as a polynomial of $t$ given below~\cite{Banerjee:2024rbc}

\begin{equation}\label{eq:HK_exp}
    \mathcal{K}(t,x,y,\widetilde{\Delta})=\frac{1}{(4\pi t)^{d/2}} e^{\frac{(x-y)^2}{4t}}e^{-M^2 t} \sum_{n=0}^\infty\frac{(-t)^n}{n!}\tilde b_n .
\end{equation}
The $\tilde{b}_n$ is the Generalized Heat-Kernel Coefficients (g-HKC).
In this context, the Green’s function is constructed using the full Heat Kernel, which captures all interactions in the Lagrangian.  
Compared to the conventional Feynman diagram approach, this method simplifies calculations by concentrating on vacuum diagrams, which reduces the number of diagrams that need to be evaluated.
The scalar interacting propagator can be presented in terms of the HK and further, be expressed by the HKCs as~\cite{Banerjee:2024rbc}
\begin{align}\label{eq:propagator}
    G(x,y)=\int_0^\infty dt\ \mathcal{K}(t,x,y,\widetilde{\Delta})=\sum_{n=0}^\infty g_n(x,y)\tilde{b}_n (x,y) .
\end{align}
Using Eqs.~\eqref{eq:HK_exp} and~\eqref{eq:propagator}, 
with \(\tilde{b}_n\) being g-HKC, the component Green's functions (CGFs) are given by 
\begin{align}\label{eq:bessel}
    g_n(x,y) &= \int_0^\infty \! \! dt \frac{1}{(4\pi t)^{\frac{d}{2}}} \, e^{\frac{z^2}{4t}} \, e^{-M^2 t}\frac{(-t)^n}{n!} 
    = \frac{(-1)^n 2^{\frac{d}{2}-n}}{(4\pi)^{\frac{d}{2}} n!} \left(\frac{M}{z}\right)^{\frac{d}{2}-n-1} \! \! \! \! \! \! \! \!  K_{\frac{d}{2}-n-1}(Mz),
\end{align}
where \(z^2=-|x-y|^2\) and \(K_n(z)\) are modified Bessel functions of the second kind. 
This approach simplifies the process by focusing on the singular behavior of the CGFs themselves. Note that the g-HKCs remain finite in the coincidence limit, \textit{i.e.,} \( x \to y \).  Only by analytical continuation in $d$ from $d<2$ to a higher dimension, we encounter the singularities of the CGFs. 
Considering the poles and the finite parts coming from the distinct two-loop diagrams and adding them, we get the total contribution in terms of the g-HKCs.
\subsection{Sunset diagram}
\begin{figure}
    \centering
   \includegraphics[width=4cm]{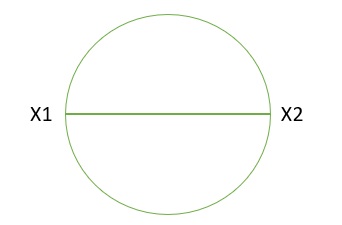}
    \includegraphics[width=5.4cm]{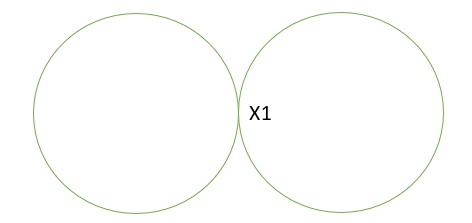}
    \includegraphics[width=2.5cm]{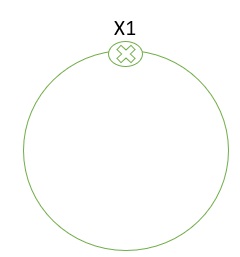}
    \caption{Left: Sunset Diagram, Middle: Infinity Diagram, Right: Counter-term Diagram}
    \label{fig:twoloop}
\end{figure}

The sunset diagram (see Fig.\,\ref{fig:twoloop}), contains the three-point vertex factor  \(V_{(3)}(x)\) defined in Eq.~\eqref{eq:vertex} and propagators. The contribution to the effective action \footnote{ The ``tr" implies only the trace over the internal symmetry indices.}, associated with the diagram is~\cite{Banerjee:2024rbc}
\begin{equation} \label{eq:l2a}
    \begin{split}
      \int d^d x \mathcal{L}_{(2)}^a \subset -\frac{1}{12} \text{tr} \Big[ \int d^d x d^d y V_{(3)}(x) G(x,y)^3 V_{(3)}(y) \Big].
    \end{split}
\end{equation}
Here, $d=4-\epsilon$. Later, we take $\epsilon \rightarrow 0$ to separate the divergent and finite terms. After expanding the two-point Green's function \(G(x, y)\) using the CGFs given in Eqs.~\eqref{app:cgf}, the contribution to the effective action reads as~\cite{Banerjee:2024rbc}
\begin{eqnarray}\label{eq:sunset}
     \int d^d x \mathcal{L}_{(2)}^a \! &=& -\frac{1}{12} \text{tr} \Big[ \int d^d x d^d y V_{(3)}(x) \Big(g_0 (x, y)^3 \Tilde{b}_0(x, y)^3 
     + 3g_0(x, y)^2 g_1(x, y)\Tilde{b}_0(x,y)^2 \Tilde{b}_1(x, y)\n\\[0.1\baselineskip]
     & +& 3\alpha g_0(x, y)^2 \Tilde{b}_0(x,y)^2 F(x,y)\Big) V_{(3)}(y) \Big],
\end{eqnarray}
where \(\alpha F(x,y) = \displaystyle\sum_{i=2} g_i \Tilde{b}_i(x,y)\) with $\alpha=\frac{1}{16\pi^2}$. We take the value of \(F\) at coincidence limit and \(\epsilon \to 0\). Note that at this limit, $F$ is a finite quantity.
From the Eq.~\eqref{eq:l2a} and using the Eq.~\eqref{eq:zinv_exp} and expanding the other CGFs up to order \(\epsilon^2\), we extract the contribution to the Lagrangian as
\begin{eqnarray} \label{l:2a}
    \mathcal{L}_{(2)}^{a} =\frac{1}{\epsilon^2}\,\mathcal{C}^{a}_{(2)}|_{\epsilon^{-2}}+ \frac{1}{\epsilon}\,\mathcal{C}^{a}_{(2)}|_{\epsilon^{-1}} + \mathcal{C}^{a}_{(2)}|_{\epsilon^{0}},
\end{eqnarray}
where the expressions of $\mathcal{C}^{a}_{(2)}|_{\epsilon^{-2}}$, $\mathcal{C}^{a}_{(2)}|_{\epsilon^{-1}}$, and $\mathcal{C}^{a}_{(2)}|_{\epsilon^{0}}$ are given in Eqs.~\eqref{app:a1}-\eqref{app:a2}.
Note that an additional finite contribution, coming from the expansion of $g_0^2(x,y)$ is given in the App.~\ref{app:finite}.
\subsection{Infinity diagram}
The infinity diagram (see Fig.\,\ref{fig:twoloop}) has the four-point vertex factor \(V_{(4)}(x)\), defined in Eq.~\eqref{eq:vertex} and propagators that are evaluated in coincidence limit, which is
\begin{eqnarray}
G(x, x)= g_0(x,x) \tilde{b}_0(x,x) + g_1(x,x) \tilde{b}_1(x,x)+\alpha F.
\end{eqnarray}
The contribution to the effective action coming from this diagram is given by~\cite{Banerjee:2024rbc} 
\begin{eqnarray} \label{eq:l2b}
 \int d^dx \mathcal{L}_{(2)}^b \subset \frac{1}{8} \text{tr} \Big[\int d^dx V_{(4)}(x)\Big(g_0(x,x) \tilde{b}_0(x,x) + g_1(x,x)\tilde{b}_1(x,x) + \alpha F\Big)^2 \Big],
\end{eqnarray}
where \(\alpha F\) is specified in the previous subsection.
Since the integral is evaluated at the coincidence limit, the CGFs can be written as
\begin{eqnarray} \label{eq:co_gn}
g_0 =  \alpha\pi^{\frac{\epsilon}{2}}2^\epsilon M^{2-\epsilon}\Gamma\left(\frac{\epsilon}{2}-1\right),\quad
g_1 = -\alpha\pi^{\frac{\epsilon}{2}}2^\epsilon M^{-\epsilon}\Gamma\left(\frac{\epsilon}{2}\right).
\end{eqnarray}
Now, after putting the values of \(g_0\) and \(g_1\) in Eq.~\eqref{eq:l2b} and expanding the other CGFs up to order \(\epsilon^2\), we extract the contribution to the Lagrangian as
\begin{eqnarray} \label{l:2b}
    \mathcal{L}_{(2)}^{b} =\frac{1}{\epsilon^2}\,\mathcal{C}^{b}_{(2)}|_{\epsilon^{-2}}+ \frac{1}{\epsilon}\,\mathcal{C}^{b}_{(2)}|_{\epsilon^{-1}} + \mathcal{C}^{b}_{(2)}|_{\epsilon^{0}},
\end{eqnarray}
where the expressions of $\mathcal{C}^{b}_{(2)}|_{\epsilon^{-2}}$, $\mathcal{C}^{b}_{(2)}|_{\epsilon^{-1}}$, and $\mathcal{C}^{b}_{(2)}|_{\epsilon^{0}}$ are given in Eqs.~\eqref{app:b1}-\eqref{app:b2}.
\subsection{Counter-term diagram}
The two-loop counter-term diagram (see Fig.\,\ref{fig:twoloop}) encapsulates the one-loop correction of the one-loop counter term. Similar to the previous two vacuum diagrams, this diagram also offers divergences up to a double pole in $\epsilon$. It is important to note that divergences in the form of $\frac{1}{\epsilon^2}$ and $\frac{1}{\epsilon}$ cancel the UV divergences, where $\frac{1}{\epsilon}\log(M^2/\mu^2)$ divergences are a signature of IR ones in the limit $M \to 0$. We check that after adding all the divergences arising from all the diagrams, the IR divergence vanishes, which validates that we have considered all the relevant contributions suitably. It contains a single propagator and one vertex factor which can be derived from the Lagrangian at one loop level~\cite{Banerjee:2023iiv}
\begin{eqnarray}\label{eq:counLag}
\mathcal{L}_{(1)} = \alpha c_s M^{-\epsilon} (4\pi)^{\frac{\epsilon}{2}} \, \text{tr} \left[  \Gamma\left(\frac{\epsilon}{2}-2\right)M^4 \Tilde{b}_0 -\Gamma\left(\frac{\epsilon}{2}-1\right)M^2 \Tilde{b}_1 + \frac{1}{2}\Gamma\left(\frac{\epsilon}{2}\right)\Tilde{b}_2  \right],    
\end{eqnarray}
where the combinatorial factor \(c_s=\frac{1}{2}, 1\) for real and complex scalar field, respectively. The vertex factor of the counter-term diagram is defined using the following equation
\begin{eqnarray}\label{eq:vct}
V_{(2)}^{(ct-1)}(x)= \frac{\partial^2 \mathcal{L}_{(1)}}{\partial\phi^2}.    
\end{eqnarray}
So, the contribution of the counter-term diagram to the effective action is ~\cite{Banerjee:2024rbc}
\begin{eqnarray} \label{eq:l2ct}
\int d^dx \mathcal{L}_{(2)}^{ct} \subset \frac{1}{2} \text{tr} \left[ \int d^dx V_{(2)}^{(ct-1)}(x) G(x,x) \right] .
\end{eqnarray}
Since this integration will be evaluated in coincidence limit, the relevant expressions of \(g_n\)s given in Eqs.~\eqref{app:cgf}, and~\eqref{app:greenfnc}, are used here. One thing to note here is that we have truncated the gamma functions in the counter-term vertex factor at the order of $1/ \epsilon$. In the counter-term vertex factor, we denote the derivatives of the HKCs with respect to the fields as  
\begin{eqnarray}\label{eq:countHeat}
 n    \tilde{b}_i^{'} = \frac{\partial \, \text{tr}( \tilde{b}_i)}{\partial \phi}, 
    \quad \tilde{b}_i^{''} = \frac{\partial^2 \, \text{tr} (\tilde{b}_i)}{\partial \phi^2}.
\end{eqnarray}
%
Because of the $1/\epsilon$ poles present in the counter-term vertex factor $V_{(2)}^{(ct-1)}$, we expand the $g_n$s up to order $\epsilon$ in Eq.~\eqref{eq:l2ct} and extract the contribution to the Lagrangian as 
\begin{eqnarray} \label{l:2ct}
    \mathcal{L}_{(2)}^{\rm ct} =\frac{1}{\epsilon^2}\,\mathcal{C}^{\rm ct}_{(2)}|_{\epsilon^{-2}}+ \frac{1}{\epsilon}\,\mathcal{C}^{\rm ct}_{(2)}|_{\epsilon^{-1}} + \mathcal{C}^{\rm ct}_{(2)}|_{\epsilon^{0}},
\end{eqnarray}
where the expressions of $\mathcal{C}^{\rm ct}_{(2)}|_{\epsilon^{-2}}$, $\mathcal{C}^{\rm ct}_{(2)}|_{\epsilon^{-1}}$, and $\mathcal{C}^{\rm ct}_{(2)}|_{\epsilon^{0}}$ are given in in Eqs.~\eqref{app:ct1}-\eqref{app:ct2}.
\subsection{Total contribution from all three diagrams}
After adding the contributions from all three diagrams given by the Eqs.~\eqref{l:2a}, \eqref{l:2b}, and \eqref{l:2ct}, we get the full two-loop correction to the Lagrangian
\begin{eqnarray} \label{l:tot}
    \mathcal{L}_{(2)} =\mathcal{L}_{(2)}^{a}+\mathcal{L}_{(2)}^{b}+\mathcal{L}_{(2)}^{\rm ct}=\frac{1}{\epsilon^2}\,\mathcal{C}_{(2)}|_{\epsilon^{-2}}+ \frac{1}{\epsilon}\,\mathcal{C}_{(2)}|_{\epsilon^{-1}} + \mathcal{C}_{(2)}|_{\epsilon^{0}},
\end{eqnarray}
%

If we substitute \(\alpha F(x,y) = \sum_{i=2}^{6} g_i \Tilde{b}_i(x,y)\) into the given expression, where the coefficients \( g_n \) are explicitly provided in Appendix~\eqref{app:greenfnc}, the Lagrangian modifies to
\begin{eqnarray}
   \mathcal{L}_{(2)}  
   = \frac{1}{\epsilon^2}\,\mathcal{C}_{(2)}|_{\epsilon^{-2}} + \frac{1}{\epsilon} \,\mathcal{C}^\prime_{(2)}|_{\epsilon^{-1}}  + \mathcal{C}^\prime_{(2)}|_{\epsilon^0}, 
\end{eqnarray}
where the expressions of $\mathcal{C}^\prime_{(2)}|_{\epsilon^{-1}}$ and  $\mathcal{C}^\prime_{(2)}|_{\epsilon^{0}}$ containing the g-HKCs are given in Eqs.~\eqref{app:F1}-\eqref{app:F2}. Note that the term \(\alpha F\) was not present in \(\mathcal{C}_{(2)}\big|_{\epsilon^{-2}}\) and so,  \(\mathcal{C}_{(2)}\big|_{\epsilon^{-2}}\) remains unchanged.
\section{Two-loop effective action in terms of \textit{U}} \label{generalU}
In this section, we consider a Lagrangian without assuming any specific form of the potential, but up to a renormalizable interaction having mass dimension four. Therefore, \(U\) (see Eq.~\eqref{eq:genlag}) is a functional of fields only and it does not contain any derivatives. For the general case, $U$ is a matrix in the field space depending on the representation of the field multiplet under symmetry. In this case, we get a generalised version of the poles in terms of \(U\) and its derivatives. For later convenience, we define $U_\phi$ and $U_{\phi\phi}$ as the derivatives of $(U)$ w.r.t the field $\phi$ as
\begin{eqnarray} \label{eq:uphi}
    (U_{\phi})^i_{jk} \equiv \frac{\partial  U_{jk}  }{\partial\phi_i} = \frac{\partial^3 V}{\partial \phi_i \partial \phi_j \partial \phi_k}, \, (U_{\phi\phi})^{ij}_{kl} \equiv \frac{\partial^2 U_{kl}  }{\partial\phi_i \partial \phi_j}= \frac{\partial^4 V}{\partial \phi_i \partial \phi_j \partial \phi_k \partial \phi_l},  \text{\hspace{0.3cm}} \forall \text{\hspace{0.3cm}} i,j,k,l=\{1,\cdots, n\} . \n \\
\end{eqnarray}

Thus, $ U_{\phi}$ is rank-[2,1] tensor and $U_{\phi\phi}$ is a rank-[2,2] tensor in field space.
To compute the vacuum diagrams we require the $n$-point vertex factors derived from the Lagrangian, which is
\begin{eqnarray}\label{eq:vertex}
V_{(n)} (x) = \frac{\partial^n \mathcal{L}}{\partial \phi^n}  .
\end{eqnarray} 
Considering $U$ has the lowest operator dimension of one, all possible g-HKCs that could contribute up to dimension six can be written as follows~\cite{Banerjee:2023iiv}: 
\begin{eqnarray}\label{eq:all_hkc}
  \tilde{b}_0(x,x)&=&I, \quad \tilde{b}_1(x,x)=U , \quad \tilde{b}_2(x,x)= U^2 + \frac{1}{3}U_{\mu\mu 
} + \frac{1}{6}G_{\mu\nu}G^{\mu\nu} , \nonumber    \\
    \tilde{b}_3(x,x)&=&\frac{1}{15} G_{\mu \nu } G_{\nu \rho } G_{\rho \mu }+\frac{1}{2} U G_{\mu \nu } G_{\mu \nu }-
    \frac{J_{\nu }^2}{10}+U^3-\frac{U_{\mu }^2}{2}, \nonumber \\
      \tilde{b}_4(x,x)&=&\frac{4}{5} U^2 G_{\mu \nu }^2-\frac{2}{5} U J_{\nu } U_{\nu }+U^4+U^2 U_{\mu \mu }+\frac{U_{\mu \mu }^2}{5}+
      \frac{(U G_{\mu \nu })^2}{5} , \\[0.35\baselineskip]
       \tilde{b}_5(x,x)&=&U^5+2U^3 U_{\mu }^2+U^2 U_{\mu \mu },\quad
    \tilde{b}_6(x,x)=U^6. \nonumber
\end{eqnarray}
The following notations are used in the expressions for the g-HKCs provided in the equation above.
\begin{equation}
\begin{split}
    & U_{\mu\mu}\equiv D^2U= - \left[P_\mu\left[P_\mu , U\right]\right], \quad U_\mu\equiv D_\mu U= - i\left[P_\mu, U\right],\\
    & G_{\mu\nu}\equiv [D_\mu , D_\nu]= - [P_\mu , P_\nu] , \quad J_\nu \equiv D_\mu G_{\mu\nu} = i \left[P_\mu [P_\mu , P_\nu]\right].
    \end{split}
\end{equation}
Here, \(P_\mu = i D_\mu\) where $D_\mu$ is the Euclidean covariant derivative. In Eq.\!~\eqref{eq:all_hkc}, the operator $U_{\mu \mu}$ in  \( \tilde{b}_2 \)  is a total derivative term. This is essentially a boundary term that goes to zero after imposing the boundary condition, so it does not contribute to the counter-term vertex factor coming from the Lagrangian at the one-loop level. It becomes significant for computing effective action if we go beyond one-loop corrections, in which the g-HKCs are located between multiple vertex factors.
\subsection{Contribution from individual diagrams}
In this subsection, we write the two-loop effective Lagrangian in terms of $U,\, U_\phi$, and $U_{\phi \phi}$ for the three individual diagrams.
\subsection*{Sunset diagram:}
From the Lagrangian given in Eq.~\eqref{l:2a}, for the sunset diagram, the coefficients of $1/\epsilon^2$, $1/\epsilon$, and the finite parts are given in terms of $U,\, U_\phi$, and $U_{\phi \phi}$ as
\begin{eqnarray}
    \mathcal{C}_{(2)}^a|_{\epsilon^{-2}}\label{eq:sunmi2}
    & =&  \frac{\alpha ^2 }{2} \text{tr} \Big[ \left(M^2+U\right) U^2_\phi \Big],\\
    \mathcal{C}_{(2)}^a|_{\epsilon^{-1}}\label{eq:sunmi1}
    & =& -\frac{\alpha ^2}{4}  \text{tr} \bigg[ U^2_\phi \left(2 F-3 M^2- U+2 \left(M^2+U\right) \log (\frac{M^2}{4\pi e^{-\gamma}})\right) + \frac{\alpha^2}{24}U_{\phi} D^2 U_{\phi} \bigg], \nonumber \\
    \mathcal{C}_{(2)}^a|_{\epsilon^{0}}\,\,\,\label{eq:sun0}
    & =& \frac{\alpha^2}{96} \text{tr} \bigg[ \Big(13+4\log\left(4\pi e^{-\gamma}\right)\Big) U_{\phi} D^2U_{\phi} + \frac{1}{96} \alpha ^2 U^2_\phi \Big[24 F (\gamma -2-\log (4 \pi ))\n\\
    &+&2 \Big[12 (\gamma -3) \gamma +\pi ^2+30\Big] M^2+24 \Big\{\log (4 \pi ) \Big((3-2 \gamma ) M^2-2 \left(M^2+U\right) \log (M^2)\n\\
    &-&2 \gamma  U 
     +U\Big)+ \log (M^2) \Big((2 \gamma -3) M^2+\left(M^2+U\right) \log (M)+2 (\gamma -1) U\Big)\n\\
     &+&\log ^2(4 \pi )  \left(M^2+U\right)\Big\}+2 \left(12 (\gamma -1) \gamma +\pi ^2-6\right) U\Big] \bigg].
\end{eqnarray}
%
\subsection*{Infinity diagram:}
From the Lagrangian given in Eq.~\eqref{l:2b}, for the infinity diagram, the coefficients of $1/\epsilon^2$, $1/\epsilon$, and the finite parts are given in terms of $U,\, U_\phi$, and $U_{\phi \phi}$ as
\begin{eqnarray}
    \mathcal{C}_{(2)}^b|_{\epsilon^{-2}}\label{eq:infmi2}
   & =&\frac{\alpha^2}{2}  \text{tr} \bigg[ \left( M^2 + U \right)^2 U_{\phi \phi } \bigg],\\
\mathcal{C}_{(2)}^b|_{\epsilon^{-1}}\label{eq:infmi1}
& =& -\frac{\alpha^2}{2}  \text{tr} \bigg[ \left(M^2+U\right) U_{\phi \phi } \left(F- M^2+\left(M^2+U\right) \log (\frac{M^2}{4 \pi e^{-\gamma}})\right) \bigg],\\
\mathcal{C}_{(2)}^b|_{\epsilon^0}\label{eq:inf0}\,\,\,
& =& \frac{\alpha^2}{48}  \text{tr} \Bigg[  U_{\phi \phi } \bigg[6 \bigg\{F- M^2+\left(M^2+U\right) \log (\frac{M^2}{4\pi e^{-\gamma}})\bigg\}^2+\left(M^2+U\right) \n\\
& \times& \bigg\{M^2 \Big(6 \gamma  (\gamma -2-2 \log (4 \pi ))+\pi ^2+6 [2+4 \log ^2(2)+\log ^2(\pi )+\log (16) \n\\
& +&\{2+\log (16)\} \log (\pi )]\Big)+12 \log (M^2) \Big(M^2 \{\gamma -1-\log (4 \pi )\}+U \{\gamma \n\\
& -&\log (4 \pi )\}\Big)+24 \left(M^2+U\right) \log ^2(M)+U \left(6 (\gamma -\log (4 \pi ))^2+\pi ^2\right)\bigg\}\bigg] \Bigg].
\end{eqnarray}

%
\subsection*{Counter-term diagram:}
From the Lagrangian given in Eq.~\eqref{l:2ct}, for the counter-term diagram, the coefficients of $1/\epsilon^2$, $1/\epsilon$, and the finite parts are given in terms of $U,\, U_\phi$, and $U_{\phi \phi}$ as
\begin{eqnarray}
    \mathcal{C}_{(2)}^{\rm ct}|_{\epsilon^{-2}} \label{eq:ctmi2}
    & = & -\alpha^2 \, \text{tr} \bigg[ \left(M^2+U\right) \Big(\left(M^2+U\right) U_{\phi \phi }+U^2_\phi  \Big) \bigg],\\
    \mathcal{C}_{(2)}^{\rm ct}|_{\epsilon^{-1}}\label{eq:ctmi1}
    & =& \frac{\alpha^2}{2}  \, \text{tr} \bigg[ \Big(\left(M^2+U\right) U_{\phi \phi }+ U^2_\phi   \Big) \left(F- M^2+\left(M^2+U\right) \log (\frac{M^2}{4\pi e^{-\gamma}})\right) \bigg], \nonumber \\
    \mathcal{C}_{(2)}^{\rm ct}|_{\epsilon^0}\label{eq:ct0}\,\,\,
    & = & -\frac{\alpha^2}{48}  \, \text{tr} \bigg[ \bigg\{M^2 \Big\{6 \gamma \Big(\gamma -2-2 \log (4 \pi )\Big)+\pi ^2+6 \Big(2+4 \log ^2(2)+\log ^2(\pi ) \n\\
    & + & \log (16) + \big( 2+\log (16)\big) \log (\pi )\Big)\Big\} + 12 \log (M^2) \Big\{M^2 \big(\gamma -1 -\log (4 \pi )\big)\n\\
    & + & U \big(\gamma -\log (4 \pi )\big)\Big\} + 24 \left(M^2+U\right) \log ^2(M)+U \Big(6 (\gamma -\log (4 \pi ))^2+\pi ^2\Big)\bigg\}\n \\
    & \times& \Big(\left(M^2+U\right) U_{\phi \phi } + U^2_\phi  \Big) \bigg].
\end{eqnarray}
\subsection{Resultant contribution}
After adding up all the contributions from all three diagrams 
, we get the total contribution for the two-loop effective action.
The coefficients of \(1/\epsilon^2\), \(1/\epsilon\), and the finite parts mentioned in Eq.~\eqref{l:tot},  are given in terms of $U,\, U_\phi$, and $U_{\phi \phi}$ as
\begin{eqnarray} 
    \mathcal{C}_{(2)}|_{\epsilon^{-2}} \label{eq:tot2}
    & =& -\frac{\alpha^2}{2} \, \text{tr} \bigg[ \left(M^2+U\right) \Big(\left(M^2+U\right) U_{\phi \phi }+ U^2_\phi   \Big) \bigg],\\
    \mathcal{C}_{(2)}|_{\epsilon^{-1}}\label{eq:tot1}
    & =& \frac{\alpha^2}{24} \, \text{tr} \Big[ U_{\phi} D^2 U_{\phi} + 6 U^2_\phi (U+M^2)\Big] ,\\
   \mathcal{C}_{(2)}|_{\epsilon^0}\,\,\,\label{eq:tot0}
   & =& \frac{\alpha^2}{96} \, \text{tr} \Bigg[ \Big\{13+4\log(4\pi e^{-\gamma})\Big\} U_{\phi} D^2U_{\phi} + U^2_\phi \bigg\{12 \bigg(2 F (\gamma -2-\log (4 \pi ))\n\\
   & +& M^2 \Big\{\gamma ^2-2 \gamma  (2+\log (4\pi ))+3+4 \log ^2(2)+\log (256)+\log (\pi ) [4+\log (16\pi )]\Big\}\n\\
   & +& U \Big\{\gamma ^2-2 \gamma  (1+\log (4\pi ))-1-4 \log ^2(2)+\log (\pi ) [2+\log (\pi )]+\log (16)[1  \n\\
   & +& \log (4\pi )]\Big\}\bigg)+48 (\gamma -2-\log (4 \pi )) \left(M^2+U\right) \log (M)\bigg\} +12 U_{\phi \phi } \bigg\{F- M^2 \n\\
   & +& \left(M^2+U\right) \log (\frac{M^2}{4\pi e^{-\gamma}})\bigg\}^2 \,\Bigg] .
\end{eqnarray}
Here, the operators $U_{\phi}$ and $U_{\phi \phi}$ in the Eqs.~\eqref{eq:sunmi2}-\eqref{eq:tot0} are defined in Eq.~\eqref{eq:uphi}.
%
\section{The Two-loop Lagrangian}
Using Eqs.~\eqref{eq:tot2}-\eqref{eq:tot0}, which include the poles and finite part from the three vacuum diagrams required for the two-loop effective action up to dimension six, the Lagrangian takes a more convenient form, which can be organized in the following manner: 
\begin{eqnarray}\label{eq:effLag}
    \mathcal{L} \supset \alpha^2 \, \text{tr} \bigg( \mathcal{C}_4M^4 +\mathcal{C}_2 M^2 +\mathcal{C}_0 M^0 + \mathcal{C}_{-2}M^{-2} \bigg),
\end{eqnarray}
with 
\begin{align} 
   \mathcal{C}_4 \,\, & \label{eq:wilson+4}
    =  - \dfrac{U_{\phi \phi }}{2\epsilon^2} + 
    \mathcal{C}^{[\![U_{\phi \phi }]\!]}_4 \, U_{\phi \phi }, \\[5pt]
  \mathcal{C}_2\,\, & 
  =  -\frac{ \left(U^2_\phi  +2 U U_{\phi \phi }\right)}{2 \epsilon ^2} + \frac{ U^2_\phi  }{4 \epsilon } 
     + \mathcal{C}^{[\![U^2_{\phi}]\!]}_2 \, U^2_\phi  + \mathcal{C}^{[\![U U_{\phi \phi }]\!]}_2 \, U U_{\phi \phi },  \\[5pt]
   \mathcal{C}_0 \,\, & 
   = -\frac{ U \left(U^2_\phi +U U_{\phi \phi }\right)}{2 \epsilon ^2} + \frac{ U U^2_\phi }{4 \epsilon } + \frac{1}{24\epsilon}U_{\phi} D^2 U_{\phi}  + \mathcal{C}^{[\![U_{\phi } D^2 U_{\phi }]\!]}_0 \, U_{\phi } D^2 U_{\phi }
   \n \\[2pt]
    &+ \mathcal{C}^{[\![U U_{\phi}^2]\!]}_0 \, U U^2_\phi
    + \mathcal{C}^{[\![U^2 U_{\phi \phi }]\!]}_0 \, U^2 U_{\phi \phi } + \mathcal{C}^{[\![U_{\mu \mu } U_{\phi \phi }]\!]}_0 \, U_{\mu \mu } U_{\phi \phi } + \mathcal{C}^{[\![G_{\mu \nu }^2 U_{\phi \phi }]\!]}_0 \, G_{\mu \nu }^2 U_{\phi \phi }, \n\\[10pt]
   \mathcal{C}_{-2} &
    = \mathcal{C}^{[\![U^2 U_{\phi}^2]\!]}_{-2} \, U^2 U^2_\phi  + \mathcal{C}^{[\![G_{\mu \nu }^2 U^2_\phi ]\!]}_{-2} \, G_{\mu \nu }^2 U^2_\phi   + \mathcal{C}^{[\![U_{\mu \mu } U^2_\phi ]\!]}_{-2} \, U_{\mu \mu } U_{\phi }^2 + \mathcal{C}^{[\![U^3 U_{\phi \phi }]\!]}_{-2} \, U^3 U_{\phi \phi } \n \\[2pt]  
    &+ \mathcal{C}^{[\![U G_{\mu \nu }^2 U_{\phi \phi }]\!]}_{-2} \, U G_{\mu \nu }^2 U_{\phi \phi }
    + \mathcal{C}^{[\![G_{\mu \nu } G_{\nu \rho } G_{\rho \mu } U_{\phi \phi }]\!]}_{-2} \, G_{\mu \nu } G_{\nu \rho } G_{\rho \mu } U_{\phi \phi } + \mathcal{C}^{[\![J_{\nu }^2 U_{\phi \phi }]\!]}_{-2} \, J_{\nu }^2 U_{\phi \phi } \n \\[2pt]
    & + \mathcal{C}^{[\![U_{\mu }^2 U_{\phi \phi }]\!]}_{-2} \, U_{\mu }^2 U_{\phi \phi }\,+\, \mathcal{C}^{[\![U U_{\mu \mu} U_{\phi \phi }]\!]}_{-2} \, U U_{\mu\mu}U_{\phi \phi }. \label{eq:wilson-2}
\end{align} \label{coeff1}
 
Note that for a renormalizable UV theory, the operator $U_{\phi \phi}$ is essentially a unit matrix with a multiplicative factor; therefore, the operator $U_{\mu \mu} U_{\phi \phi}$ turns out to be a total derivative term. All the coefficients of the operators such as $U_{\phi \phi}$, $U_{\phi}^2$, $U_{\phi} D^2 U_{\phi}$ \textit{etc.} in the above equations are given below.
\begin{align}
    &\mathcal{C}^{[\![U_{\phi \phi }]\!]}_4 = \frac{1}{8}  \left[1-\log(\frac{M^2}{4 \pi  e^{-\gamma}})\right]^2,\qquad 
    \mathcal{C}^{[\![U_{\phi } D^2 U_{\phi }]\!]}_0 \, = \frac{1}{96} \Big[13 +4 \log (4 \pi e^{-\gamma})\Big],\n \\[0.1\baselineskip]
    &\mathcal{C}^{[\![U_{\phi}^2]\!]}_2 \,\,  = \frac{1}{8}   \Big[3+\gamma ^2-2 \gamma  \big(2+\log (4\pi )\big)-2 \left(2 +\log (4 \pi e^{-\gamma} )\right) \log (M^2)\n \\[0.1\baselineskip]
    & \hspace{1cm}  +4 \log ^2(2)+\log (256)+\log (\pi ) \big(4+\log (16\pi )\big)\Big],\n \\[0.1\baselineskip]
   & \mathcal{C}^{[\![U U_{\phi \phi }]\!]}_2 \quad = - \frac{1}{4}  
      \Bigg[\left\{1- \log (\frac{M^2}{4\pi e^{-\gamma}})\right\} \log (\frac{M^2}{4\pi e^{-\gamma}})\Bigg],\n\\[0.1\baselineskip]
      & \mathcal{C}^{[\![U U_{\phi}^2]\!]}_0 \, =  - \frac{1}{8}   \Big[1-\gamma ^2+2 \gamma -2 (\gamma-2)  \log (M^2)
      -2 \log (4 \pi ) \Big\{1-\gamma - \log (M^2)\Big\} 
     - \log ^2(4 \pi )\Big], \n \\[0.1\baselineskip]
    & \mathcal{C}^{[\![U^2 U_{\phi \phi }]\!]}_0 \,\,\,\,\,\, 
    = - \frac{1}{8}  \Big[1- \log \left(\frac{M^2}{4\pi e^{-\gamma}}\right)-\log ^2\left(\frac{M^2}{4\pi e^{-\gamma}}\right)\Big],\n\\[0.1\baselineskip]
    & \mathcal{C}^{[\![U_{\mu \mu } U_{\phi \phi }]\!]}_0 \,\,\,\, = - \frac{1}{24}  \left[1-\log (\frac{M^2}{4\pi e^{-\gamma}})\right],\quad
     \mathcal{C}^{[\![G_{\mu \nu }^2 U_{\phi \phi }]\!]}_0 \,=   -  \frac{1}{48}   \left[1-\log (\frac{M^2}{4\pi e^{-\gamma}})\right],\\[0.1\baselineskip]
    &\mathcal{C}^{[\![U^2 U_{\phi}^2]\!]}_{-2} \quad\,\,\, = -\frac{1}{8} \Big[2+\log (4 \pi e^{-\gamma})\Big],\qquad\,\,\,\,
     \mathcal{C}^{[\![G_{\mu \nu }^2 U_{\phi}2]\!]}_{-2} \, =- \frac{1}{48}  \Big[2+\log (4 \pi e^{-\gamma})\Big],\n\\[0.1\baselineskip]
    & \mathcal{C}^{[\![U_{\mu \mu } U_{\phi}^2]\!]}_{-2} \quad\, =- \frac{1}{24}  
     \Big[2+\log (4 \pi e^{-\gamma})\Big], \qquad
    \mathcal{C}^{[\![U^3 U_{\phi \phi }]\!]}_{-2} \,= \frac{1}{24}  \Bigg[1+2 \log (\frac{M^2}{4\pi e^{-\gamma}}) \Bigg],\n \\[0.1\baselineskip]
 & \mathcal{C}^{[\![U G_{\mu \nu }^2 U_{\phi \phi }]\!]}_{-2} \, =  \frac{1}{48}  ,\hspace{1cm}
     \mathcal{C}^{[\![G_{\mu \nu } G_{\nu \rho } G_{\rho \mu } U_{\phi \phi }]\!]}_{-2} \, =\frac{1}{360}  \left[1-\log (\frac{M^2}{4\pi e^{-\gamma}})\right], \n\\[0.1\baselineskip]
    & \mathcal{C}^{[\![J_{\nu }^2 U_{\phi \phi }]\!]}_{-2} \quad\,\, =- \frac{1}{240}  \left[1-\log (\frac{M^2}{4\pi e^{-\gamma}})\right] ,\quad\,
     \mathcal{C}^{[\![U_{\mu }^2 U_{\phi \phi }]\!]}_{-2} \,=- \frac{1}{48}  \left[1-\log (\frac{M^2}{4\pi e^{-\gamma}})\right],\n\\[0.1\baselineskip]
     & \mathcal{C}^{[\![U U_{\mu \mu } U_{\phi \phi }]\!]}_{-2} \, =  \frac{1}{24} \log\left(\frac{M^2}{4\pi e^{-\gamma}}\right).\n
\end{align}

Note that in $MS$ regularization scheme, the log$(M^2)$ reads as $\log\left(M^2/\mu^2\right)$, whereas in $\overline{MS}$ regularization scheme $\mu^2$ rescales as $\mu^2\to\mu^2/ (4\pi e^{-\gamma})$, where $\mu$ is renormalization scale of the theory.
For some UV scenarios, the mass dimension of the operators, contained in $U$, can be one, e.g., $U = \kappa \phi$ with $\kappa$ as a dimension-full coupling. In that case, to compute all the dimension six operators, we have to consider terms up to $\mathcal{C}_{-10}$. To encapsulate such scenarios, we have provided the additional contributions necessary, i.e. $\mathcal{C}_{-4}, \mathcal{C}_{-6}, \mathcal{C}_{-8},\mathcal{C}_{-10}$ in App.~\ref{app:coeff}. 

At this point it is worth to mention that the algebraic form of the effective Lagrangian up to dimension six in Eq.~\ref{eq:effLag} is independent of any specific UV scenarios that contain a heavy scalar that is going to be integrated out. The important and useful part of this construction is that one can easily compute the effective operators along with the WCs even without knowing the underlying HK method that has been employed to get this form of the Lagrangian. In the following sections, for two example models, we have elaborated on this in detail.

  Note that only the finite parts contribute to the effective action, whereas the divergences help us to compute the beta function, coming from the necessary counterterms that need to be added to cancel the UV divergences. This work leaves out the divergences; however, an extensive analysis of the evolution of the renormalization group (RG) is provided in~\cite{Banerjee:2024rbc}.
\begin{table}[ht]\label{bosonic_ops}
\centering
\begin{tabular}{|| rcl || rcl ||}\hline
 \(\scO_{GG}\) &\(=\)& \(g_s^2 \abs{H}^2G_{\mu \nu }^a G^{a, \mu \nu }\) & \(\scO_H\)   &\(=\)& \(\frac{1}{2}\big(\pd_{\mu} \abs{H}^2\big)^2\)\\
 \hline
 \(\scO_{WW}\) &\(=\)& \(g_W^2  \abs{H}^2 W_{\mu \nu }^aW^{a,\mu \nu } \) &  \(\scO_T\)   &\(=\)& \(\frac{1}{2}\big( H^{\dag} \Dfbd H\big)^2\) \\
 \hline
 \(\scO_{BB}\) &\(=\)& \(g_Y^2 \abs{H}^2 B_{\mu \nu }B^{\mu \nu }\) & \(\scO_R\)   &\(=\)& \(\abs{H}^2\abs{D_{\mu}H}^2\) \\
 \hline
 \(\scO_{WB}\) &\(=\)& \(2g_W g_Y H^\dag {\tau^a}H W_{\mu \nu }^a B^{\mu \nu }\) &  \(\scO_D\)   &\(=\)& \(\abs{D^2H}^2\) \\
 \hline
 \(\scO_W\)   &\(=\)& \(ig_W\big(H^\dag \tau^a \Dfb H\big)D^\nu W_{\mu \nu }^a\) &  \(\scO_6\)   &\(=\)& \(\abs{H}^6\) \\
 \hline
 \(\scO_B\)   &\(=\)& \(ig_Y Y\big(H^\dag \Dfb H\big)\pd^\nu B_{\mu \nu }\)  &  \(\scO_{2G}\) &\(=\)& \(-\frac{1}{2} \big(D^\mu G_{\mu \nu }^a\big)^2\) \\
 \hline
 \(\scO_{3G}\) &\(=\)& \(\frac{1}{3!}g_sf^{abc}G_\rho ^{a\mu }G_\mu ^{b\nu }G_\nu ^{c\rho }\) &  \(\scO_{2W}\) &\(=\)& \(-\frac{1}{2} \big(D^\mu W_{\mu \nu }^a\big)^2\) \\
 \hline
 \(\scO_{3W}\) &\(=\)& \(\frac{1}{3!}g_W \epsilon^{abc}W_\rho ^{a\mu }W_\mu ^{b\nu }W_\nu ^{c\rho }\) & \(\scO_{2B}\) &\(=\)& \(-\frac{1}{2} \big(\pd^{\mu} B_{\mu \nu }\big)^2\) \\
  \hline
\end{tabular}
\caption{CP-conserving dimension six bosonic operators in SILH basis~\cite{Hagiwara:1993ck,Giudice:2007fh,Henning:2014wua}.} \label{tbl:operators}
\vspace{-10pt}
\end{table}
\renewcommand\arraystretch{0}
\section{Two-loop effective action for triplet and doublet models} \label{models}
Here, we will briefly discuss the local expansion of the heavy field at its classical value. This expansion ensures that the effective action can be written in a manifestly gauge-invariant way. 
The local effective action can be computed after incorporating the covariant derivative expanded form of the classical solution of the heavy field.

We will discuss this in detail for the specific model examples in subsequent sections.  We consider a generic form of the Lagrangian~\cite{Henning:2014wua,Gaillard:1985uh,Cheyette:1987qz},
\begin{equation}
\mathcal{L}[\Phi,\phi] \supset  -\Phi^{\dag}\big(D^2 + m^2 + U(x) \big) \Phi + \big(\Phi^{\dag}B(x) + \text{h.c.}\big),
\label{eqn:tree_UV_Lag}
\end{equation}
where $\Phi$ and $\phi$ are the heavy and light scalar fields, respectively. Here, \(B(x)\) is a generic function of the light fields $\phi(x)$ whereas $U(x)$ can be a generic function of both the heavy fields $\Phi(x)$ and the light fields $\phi(x)$.
The equation of motion (EOM) for \(\Phi\) is 
\begin{equation}
\big(D^2 + m^2 + U(x)\big) \Phi = B(x).
\end{equation}
After linearizing, \textit{i.e.}, ignoring the $\mathcal{O}(\Phi^2)$ terms in the EOM and solving it in the case where $p^2 \ll m^2$, as prescribed in~\cite{Henning:2014wua}, we get
\begin{align}
\Phi_c 
&= \frac{1}{m^2} B - \frac{1}{m^2}\big(D^2 + U\big)\frac{1}{m^2}B + \frac{1}{m^2}\big(D^2 + U\big)\frac{1}{m^2}\big(D^2 + U\big)\frac{1}{m^2}B + \dots \hspace{2mm}.
\label{eqn:phic_expand}
\end{align}
The mass-squared matrix in the equation above need not be diagonal, so $1/m^2$ may not commute with $U$. Back into the Lagrangian, we plug \(\Phi_c\) to obtain the tree-level effective action. Note that even though we only considered $U(x)$ to be only a function of $\phi$, we can include $\Phi$ as well and use recursion to get additional correction terms in the classical solution of $\Phi$.
%
%

As an example scenario for computing the two-loop effective action, we take into consideration two particular models: the two Higgs doublet model (2HDM), the extension of the SM by an extra Higgs doublet $\Phi$ carrying hypercharge $Y_\Phi=-1/2$, and the complex triplet model, the extension of the SM by an electroweak scalar triplet $\Delta$ with hypercharge $Y_\Delta=1$. These two well-known models have significant phenomenological implications, as discussed in Sec.\,\ref{sec:intro}. We provide our calculation procedures and results for these two cases in the following subsections. 
\subsection{Electroweak triplet with hypercharge \texorpdfstring{$Y_\Delta=1$}{Lg}}
In this subsection, we integrate out an additional electroweak scalar $\Delta$ that transforms as a triplet under the SM gauge group, to get two-loop effective action. The triplet scalar's mass ($m_\Delta$) is significantly higher than the electroweak symmetry-breaking scale\footnote{$v\approx 246$ GeV is the \textit{vev} of the Higgs field.} \textit{i.e.,} $m_\Delta\gg v$. When the triplet scalar $\Delta$ interacts with the SM Higgs doublet $(H)$ and leptons, the most general UV Lagrangian  can be expressed as~\cite{Barrie:2022cub, Bakshi:2018ics, arhrib:hal-00608687,Schechter:1981cv,Schechter:1980gr,FileviezPerez:2008jbu,Das:2023zby}
\begin{equation}
\label{lbsmCTS}
    \mathcal{L} = \mathcal{L}_{\rm SM} \ + \text{tr} [ (D_{\mu} \Delta)^\dagger (D^{\mu} \Delta ) ] - m_{\Delta}^{2} \text{tr} [ \Delta^{\dagger} \Delta ] - \mathcal{L}_{Y} - V(H,\Delta) \, ,
\end{equation}
where,
\begin{eqnarray}\label{eq:trptpot}
	    V(H,\Delta) &=&\,  \lambda_{1} \big(H^{\dagger} H\big) \text{tr}[ \Delta^{\dagger} \Delta ] + \lambda_{2} \big(\text{tr}[ \Delta^{\dagger} \Delta ]\big)^2 + \lambda_{3}\,\text{tr}\big[ \big(\Delta^{\dagger} \Delta\big)^2 \big]  \\
	    & +&\, \lambda_{4}\big( H^\dagger \Delta {\Delta}^\dagger H \big) + \big[ \mu_{\Delta}\big( H^{T} i \sigma_2 \Delta^{\dagger} H \big) + \text{h.c.}\big] \, , 
\end{eqnarray}
and,
\vspace{-3mm}
\begin{eqnarray} \label{lep1}
    \mathcal{L}_{Y} = \, Y_{\Delta} l_{L}^{T} C i \sigma_2 \Delta \, l_{L} + \text{h.c.} ~~~~~~~~~~~~~~~~~~~~
\end{eqnarray}
Here, $\Delta = \Delta^a \tau^a$, where $\tau^a = \frac{\sigma^a}{2}$, are the generators in the fundamental representation of $SU(2)$. The covariant derivative in the kinetic term, $D_{\mu}=(\partial_{\mu} - i g_W W_{\mu}^a \tau^a - i g_Y Y B_{\mu})$, where
the gauge fields $W_\mu^a$ are in the adjoint representation of the $SU(2)_L$ group.
Each parameter in the potential above is treated as a real parameter, whereas the $Y_\Delta$ in $\mathcal{L}_{Y}$  is generally a complex parameter.

To compute the two-loop contributions, we start with Eqs.~\eqref{eq:effLag}–\eqref{eq:wilson-2}, from which it is evident that only the $U$ matrix and its functional derivatives {\it w.r.to} the fields, i.e., $U_\phi$ and $U_{\phi \phi}$ are necessary. These quantities encapsulate all the relevant model parameters and fields necessary for determining the effective operators and their corresponding Wilson coefficients. In the following section,
using the Eqs.~\eqref{app:trpttrace1}-\eqref{app:trpttrace2}, we can write the potential term on this basis given below.
\begin{equation}
   \mathcal{L} \supset \dfrac{1}{2} \begin{pmatrix}
       \Delta^\ast_i & \Delta_i 
   \end{pmatrix} \, 
   U_{ij}
   \begin{pmatrix}
       \Delta_j \\ \Delta^\ast_j
   \end{pmatrix}
   = \dfrac{1}{2} \begin{pmatrix}
       \Delta^\ast_i & \Delta_i 
   \end{pmatrix} \, 
   \begin{pmatrix}
       \left(U_{11}\right)_{ij} & \left(U_{12}\right)_{ij} \\
       \left(U_{21}\right)_{ij} & \left(U_{22}\right)_{ij}
   \end{pmatrix} \,
   \begin{pmatrix}
       \Delta_j \\ \Delta^\ast_j
   \end{pmatrix},
\end{equation}
where $i,j$ runs from 1 to 3. Essentially $U$ is a $6\times6$ hermitian matrix decomposed by four $3\times 3$ sub-matrices. These four sub-matrices can be defined as
\begin{eqnarray}\label{eq:matrixtriplet}
        \left(U_{11}\right)_{ij} \equiv \dfrac{\partial^2 V}{\partial \Delta_i^\ast \partial \Delta_j},\quad
        \left(U_{22}\right)_{ij} \equiv \dfrac{\partial^2 V}{\partial \Delta_i \partial \Delta_j^\ast},\quad  \left(U_{22}\right)_{ij}= \left(U^\ast_{11}\right)_{ij},\n\\
      \left(U_{12}\right)_{ij} \equiv \dfrac{\partial^2 V}{\partial \Delta_i^\ast \partial \Delta_j^\ast},\quad
        \left(U_{21}\right)_{ij} \equiv \dfrac{\partial^2 V}{\partial \Delta_i \partial \Delta_j},\quad  \left(U_{21}\right)_{ij}= \left(U^\ast_{12}\right)_{ij}.
\end{eqnarray}
The elements of these four matrices are expressed explicitly in terms of the model parameters as
\begin{eqnarray}
        (U_{{11}})_{ij} 
         &=& \dfrac{1}{2} \lambda_1 \lvert H \rvert ^2 \delta_{ij} + \dfrac{1}{2}\lambda_2 \Big(\, \lvert \Delta \rvert ^2 \delta_{ij} + \Delta_i \Delta^*_j \, \Big) \n \\
        & +& \dfrac{1}{2} \lambda_3 \Big(\, \lvert \Delta \rvert ^2 \delta_{ij} + \Delta_i \Delta^*_j - \Delta^* _i \Delta_j \, \Big)
        + \lambda_4 \left(\dfrac{1}{4} \lvert H \rvert ^2 \delta_{ij} - \frac{i}{2} \epsilon^{ijk} ( H^\dagger \tau^k H) \right), \\
        \n \\
        (U_{{12}})_{ij} & =& \dfrac{1}{2} \lambda_2  \Delta_i \Delta_j + \dfrac{1}{4}\lambda_3 \Big(2 \Delta_i \Delta_j -\Delta_k \Delta_k \delta_{ij}\Big).
\end{eqnarray}
Using the CDE method, we can write the solution (see Eq.~\eqref{eqn:phic_expand}) for the classical background field, up to linear approximation as
\begin{eqnarray}
    (\Delta_c)^i &=& \frac{1}{m_\Delta^2}B^i + \frac{1}{m_\Delta^4} \left(p^2\delta_{ij}-\left(U_{11}\right)_{ij}\right) B^j +\mathcal{O}(m_\Delta^{-6}),
\end{eqnarray}
where $ B^i = \mu_\Delta H^T i \sigma_2 \tau^i H = - \mu_\Delta \tilde{H}^\dagger \tau^i H$ with $\tilde{H}=i\sigma_2 H^\ast$ (see Eq.~\eqref{eqn:tree_UV_Lag}). Using the above expression of $\Delta_c^i$, we get the following relation up to operator dimension six
\begin{eqnarray}
  |\Delta_c|^2=\left(\Delta_c^i\right)^\ast  \Delta_c^i \supset \frac{\mu_\Delta^2}{m_\Delta^4}\bigg[\frac{1}{2}|H|^4 
      +\frac{2}{m_\Delta^2}\mathcal{O}_H - \frac{2}{m_\Delta^2}\left(\frac{\lambda_1}{2}+\frac{\lambda_4}{4}\right)\frac{\mathcal{O}_6}{2}\bigg],\label{eq:Delsq}
\end{eqnarray}
where the dimension six bosonic operators $\mathcal{O}_H$ and $\mathcal{O}_6$ are defined in Tab.\,\ref{tbl:operators}. 
 We redo the computations of the dimension six operator structures at the one-loop level for this scenario given in the App.~\ref{app:onelooptrp}. Our computation matches with the results of~\cite{Bakshi:2018ics}.

\subsubsection{Operators contributing up to dimension six at two-loop level }
Here, we discuss the operators up to dimension six generated at the two-loop level for both bosonic and leptonic cases for the complex triplet scenario.  {Before computing the operators, we would first define the 3-pt and 4-pt vertex that we defined earlier, for our model.}
\begin{eqnarray}
 &&   \frac{\partial \, (U_{11})_{ij}}{\partial \Delta_k} = (U_\phi)_{ijk} = \frac{1}{2} \lambda_2(\Delta_k \delta_{ij} + \Delta_j \delta_{ik}) + \frac{1}{2} \lambda_3 (\Delta_k \delta_{ij} + \Delta_j \delta_{ik} - \Delta_i \delta_{jk}), \n\\ 
 &&   \frac{\partial \, (U_{11})_{ij}}{\partial \Delta^*_k} =(U^*_\phi)_{ijk} 
    = \frac{1}{2} \lambda_2(\Delta^*_k \delta_{ij} + \Delta^*_j \delta_{ik}) + \frac{1}{2} \lambda_3 (\Delta^*_k \delta_{ij} + \Delta^*_j \delta_{ik} - \Delta^*_i \delta_{jk}), \\ \n
&&     \frac{\partial^2 \left(U_{11} \right)_{ij}}{\partial \Delta_k \partial \Delta_l^\ast}=(U_{\phi \phi})_{ijkl}
     = \frac{1}{2} \lambda_2(\delta_{ij} \delta_{kl} + \delta_{ik} \delta_{jl}) + \frac{1}{2} \lambda_3 (\delta_{ij} \delta_{kl} + \delta_{ik} \delta_{jl} - \delta_{il} \delta_{jk}).
\end{eqnarray}

\subsubsection*{Bosonic operators:}

Using the classical solution for the heavy field given in Eq.~\eqref{eq:Delsq}, the pure bosonic operators\footnote{Here, $G_{\mu\nu}^2 =G_{\mu\nu}G^{\mu\nu}$ and $G_{\mu\nu}^3 =G_{\mu}^\nu G_{\nu}^\rho G_{\rho}^\mu$. } generated for this scenario are given here.
\begin{eqnarray}
 &&    \dfrac{1}{2}\text{tr} \big[U_{\phi\phi}\big]\; \label{eq:trpt1}
     =  \dfrac{1}{2} \bigg[ \frac{\partial^2 \left(U_{11} \right)_{ii}}{\partial \Delta_k \partial \Delta_k^\ast} + \frac{\partial^2 \left(U_{22} \right)_{ii}}{\partial \Delta_k \partial \Delta_k^\ast} \bigg]
     = 6\lambda_2 + \dfrac{9}{2} \lambda_3 , 
     \\[0.55\baselineskip]
 &&    \dfrac{1}{2}\text{tr}\big[U_{\phi}^2\big] \label{eq:trpt2} 
     = 3 \,\bigg[ \frac{\partial \, (U_{11})_{ij}}{\partial \Delta_l}\frac{\partial \, (U_{22})_{ji}}{\partial \Delta_l^\ast} \bigg]   \n \\
     &&\hspace{15mm}= \dfrac{3}{4} \left(6\lambda^2_2 - \lambda^2_3 + 2 \lambda_2 \lambda_3\right) |\Delta_c|^2  \n \\
      &&\hspace{15mm}\supset \frac{3\mu_\Delta^2}{4 m_\Delta^4} \left(6\lambda^2_2 - \lambda^2_3 + 2 \lambda_2 \lambda_3\right)  \left[\frac{1}{2}|H|^4+\frac{2}{m_\Delta^2}\mathcal{O}_H - \frac{1}{m_\Delta^2}\left(\frac{\lambda_1}{2}+\frac{\lambda_4}{4}\right)\mathcal{O}_6\right] \!, \n\\
       \\
 &&     \dfrac{1}{2}\text{tr}\big[U U_{\phi\phi}\big]\label{eq:trpt3}
       = \dfrac{1}{2} (2 \lambda_2 + \dfrac{3}{2} \lambda_3) \big[(U_{11})_{ii} + (U_{22})_{ii}\big]  \n \\[0.15\baselineskip]
      &&\hspace{15mm}= \left(2\lambda_2 + \dfrac{3}{2} \lambda_3\right)\left[\left(\frac{3\lambda_1}{2}+\frac{3\lambda_4}{4}\right)|H|^2+\left(2\lambda_2+\frac{3\lambda_3}{2}\right)|\Delta_c|^2\right]  \n \\[0.15\baselineskip]
      &&\hspace{15mm}\supset\left(2\lambda_2 + \dfrac{3}{2} \lambda_3\right)\bigg[\left(\frac{3\lambda_1}{2}+\frac{3\lambda_4}{4}\right)|H|^2+\left(2\lambda_2+\frac{3\lambda_3}{2}\right)\n\\[0.15\baselineskip]
      &&\hspace{15mm}\times\frac{\mu_\Delta^2}{m_\Delta^4}\bigg(\frac{1}{2}|H|^4 
      +\frac{2}{m_\Delta^2}\mathcal{O}_H - \frac{1}{m_\Delta^2}\left(\frac{\lambda_1}{2}+\frac{\lambda_4}{4}\right)\mathcal{O}_6\bigg)\bigg], \\
      [0.45\baselineskip]
  &&    \dfrac{1}{2}\text{tr}\big[UU_{\phi}^2\big]
      =  \dfrac{1}{2} \bigg[ \frac{\partial \, (U_{11})_{ij}}{\partial \Delta_k} (U_{11})_{ii'} \frac{\partial \,  (U_{11})_{i'j}}{\partial \Delta_k^\ast} + \frac{\partial \,  (U_{22})_{ij} }{\partial \Delta_k^\ast} (U_{22})_{ii'} \frac{\partial \,  (U_{22})_{i'j} }{\partial \Delta_k} \n \\[0.15\baselineskip]
      &&\hspace{15mm}+  \frac{\partial \,  (U_{11})_{ij} }{\partial \Delta_k} (U_{12})_{ii'} \frac{\partial \, (U_{11})_{i'j} }{\partial \Delta_k} + \frac{\partial \, (U_{22})_{ij} }{\partial \Delta_k^\ast} (U_{21})_{ii'} \frac{\partial \, (U_{22})_{i'j} }{\partial \Delta_k^\ast} \bigg], \n \\[0.15\baselineskip]
      &&\hspace{15mm}= \dfrac{3}{4} \left(\lambda_1+\frac{\lambda_4}{2}\right) \left(6\lambda^2_2 - \lambda^2_3 + 2 \lambda_2 \lambda_3\right) |H|^2|\Delta_c|^2 , \n\\[0.15\baselineskip]
      &&\hspace{15mm}\supset {\left(\lambda_1+\frac{\lambda_4}{2}\right) \left(6\lambda^2_2 - \lambda^2_3 + 2 \lambda_2 \lambda_3\right) \frac{3\mu_\Delta^2}{8m_\Delta^4}\mathcal{O}_6} ,
      \\[0.45\baselineskip]
  &&   \dfrac{1}{2}\text{tr}\big[U^2 U_{\phi\phi}\big]=  \dfrac{1}{2} \Big[ \frac{1}{2}( \lambda_2 + \lambda_3 )( U_{11} + U_{22})_{ij} ( U_{11} + U_{22})_{ji} + \frac{\lambda_2}{2} ( U_{11}^2 + U_{22}^2 + 2 U_{12} U_{21} )_{ii} \Big]  \n \\
     &&\hspace{15mm}= \dfrac{1}{2} \left[\left(\lambda_2 + \frac{1}{2} \lambda_3 \right)\big(U_{11}^2 + U^2_{22}\big)_{ii} + \lambda_2 \big(U_{12}\big)_{ij} \big(U_{21} \big)_{ji} + ( \lambda_2 + \lambda_3 ) \big(U_{11} \big)_{ij} \big(U_{22} \big)_{ji} \right] \n  \\
      &&\hspace{15mm}\supset \left(\dfrac{3}{2} \lambda_2 + \lambda_3 \right)\bigg[\left(2\lambda_1\lambda_2+\frac{3\lambda_1\lambda_3}{2}+\lambda_2\lambda_4+\frac{3\lambda_3\lambda_4}{4}\right)\frac{\mu_\Delta^2}{2m_\Delta^4}\mathcal{O}_6\n \\
      &&\hspace{15mm}+ \left(\frac{3\lambda_1^2}{4}+\frac{5\lambda_4^2}{16}+\frac{3\lambda_1\lambda_4}{4}\right)|H|^4\bigg], 
      \\[0.45\baselineskip]
  &&    \dfrac{1}{2}\text{tr}\big[U^3 U_{\phi\phi}\big]=
      \dfrac{1}{2} \bigg[ \left(\dfrac{3}{2} \lambda_2 + \lambda_3 \right)\big(U_{11}^3+U^3_{22} \big)_{ii}+ \lambda_2 \big(3 U_{11}U_{12}U_{21}+3 U_{22}U_{21}U_{12}\big)_{ii}\bigg]
      \n \\
      &&\hspace{15mm}\supset \frac{3}{8} \left(\dfrac{3}{2} \lambda_2 + \lambda_3 \right)\left(\lambda_1^3+\frac{3\lambda_4^3}{8}+\frac{3 \lambda_1^2\lambda_4}{2}+\frac{5\lambda_1\lambda_4^2}{4}\right)\mathcal{O}_6, 
       \\[0.45\baselineskip]
   &&   \dfrac{1}{2}\text{tr}\big[U_{\phi}D^2U_{\phi}\big] = \dfrac{1}{2} \left[\frac{\partial \, U_{ij} }{\partial \Delta_k}D^2\frac{\partial \, U_{ji} }{\partial \Delta_k^\ast}\right]\n \\
    &&\hspace{15mm}= \dfrac{3}{4} \left(6\lambda^2_2 - \lambda^2_3 + 2 \lambda_2 \lambda_3\right)  \left(\Delta_c^i\right)^\ast D^2 \Delta_c^i \n\\
     &&\hspace{15mm}\supset \dfrac{3}{4} \left(6\lambda^2_2 - \lambda^2_3 + 2 \lambda_2 \lambda_3\right) \left(-\frac{\mu_\Delta^2}{m_\Delta^4}\mathcal{O}_H\right),
     \\[0.45\baselineskip]
  &&  \dfrac{1}{2}\text{tr}\big[G_{\mu\nu}^2U_{\phi\phi}\big] 
    = \dfrac{1}{2} \left(2 \lambda_2 + \dfrac{3}{2} \lambda_3\right) \text{tr}\left[G_{\mu\nu}^2\right]\n \\
      &&\hspace{15mm}= -\left(2\lambda_2 + \dfrac{3}{2} \lambda_3\right)\left(2 g_W^2 \left(W_{\mu\nu}^a\right)^2+ 3 g_Y^2 \left(B_{\mu\nu}\right)^2 \right),
     \\[0.4\baselineskip]
  &&   \dfrac{1}{2}\text{tr}\big[UG_{\mu\nu}^2U_{\phi\phi}\big]= \dfrac{1}{2} \left(2 \lambda_2 + \dfrac{3}{2} \lambda_3\right) \text{tr}\left[UG_{\mu\nu}^2\right] \n \\ 
     &&\hspace{15mm}\supset -\left(2\lambda_2 + \dfrac{3}{2} \lambda_3\right)\left[\left(\frac{\lambda_1}{2}+\frac{\lambda_4}{4}\right)\Big(2\mathcal{O}_{WW}+3\mathcal{O}_{BB}\Big)-\lambda_4\mathcal{O}_{WB}\right], \\[0.4\baselineskip]
&&    \dfrac{1}{2} \text{tr}\big[J_{\nu}^2U_{\phi\phi}\big]
    = \dfrac{1}{2} (2 \lambda_2 + \dfrac{3}{2} \lambda_3) \, \text{tr}\left[(D_{\mu} G_{\mu \nu})^2\right] \n \\
     &&\hspace{15mm}= -\left(2\lambda_2 + \dfrac{3}{2} \lambda_3\right)\Big(4 g_W^2\mathcal{O}_{2W}+6 g_Y^2\mathcal{O}_{2B}\Big),\\[0.4\baselineskip]
 &&   \dfrac{1}{2} \text{tr}\big[U_{\mu}^2U_{\phi\phi}\big] 
    = \dfrac{1}{2} \left[\left(\dfrac{3}{2} \lambda_2 + \lambda_3 \right) \! \left((D_\mu U_{11})^2 + (D_{\mu} U_{22})^2\right)_{ii} + \lambda_2 \big(D_\mu U_{12}\big)_{ij} \big(D_\mu U_{21}\big)_{ji} \right] \n \\
     &&\hspace{15mm}\supset \dfrac{1}{2} \left(\dfrac{3}{2} \lambda_2 + \lambda_3 \right) \! \left[2\left(\frac{3\lambda_1^2}{4}+ \frac{3\lambda_4^2}{16}+\frac{3\lambda_1\lambda_4}{4}\right)\mathcal{O}_H+\frac{\lambda_4^2}{4}\left(\mathcal{O}_T+2\mathcal{O}_R\right)\right]\!,  \\[0.4\baselineskip]
 &&     \dfrac{1}{2} \text{tr}\big[U U_{\mu\mu}U_{\phi\phi}\big] =
    - \dfrac{1}{2} \left[\left(\dfrac{3}{2} \lambda_2 + \lambda_3 \right)\big((D_\mu U_{11})^2 + (D_{\mu} U_{22})^2\big)_{ii} + \lambda_2 \big(D_\mu U_{12}\big)_{ij} \big(D_\mu U_{21}\big)_{ji} \right]\n \\
     &&\hspace{15mm}\supset \! - \dfrac{1}{2} \left(\dfrac{3}{2} \lambda_2 + \lambda_3 \right) \! \left[2\left(\frac{3\lambda_1^2}{4}+ \frac{3\lambda_4^2}{16}+\frac{3\lambda_1\lambda_4}{4}\right)\mathcal{O}_H+\frac{\lambda_4^2}{4}\left(\mathcal{O}_T+2\mathcal{O}_R\right)\right] \!\!,  \\[0.4\baselineskip]
  &&   \dfrac{1}{2} \text{tr}\big[G_{\mu\nu}^3 U_{\phi\phi}\big]
     = \dfrac{1}{2} (2 \lambda_2 + \dfrac{3}{2} \lambda_3) \text{tr}\left[G_{\mu\nu}^3\right] 
       =  -\left(2\lambda_2 + \dfrac{3}{2} \lambda_3\right) 6 g_W^2 \mathcal{O}_{3W}.\label{eq:trpend}
     \end{eqnarray}
We have used the relations given in Eqs.~\eqref{app:essential1}-\eqref{app:essential2} to simplify and write the effective operators in the SILH basis operators listed in Tab.\,\ref{tbl:operators}

Note that the operators $ U_{\mu\mu} U_\phi^2$ , $U^2 U_\phi^2$ and $G_{\mu \nu}^2 U_\phi^2$ (see Eq.~\eqref{eq:wilson-2}) are excluded here as they generate operators of dimension eight or higher. 

Because $U_{\phi\phi}$ is a diagonal matrix, as mentioned previously, the operators $U U_{\mu \mu}U_{\phi \phi}$ and $U^2_\mu U_{\phi \phi}$ can be related by integration by parts (IBP) \textit{i.e.,} $\text{tr}\big[U U_{\mu\mu} U_{\phi\phi}\big]=-\text{tr}\big[U_{\mu}^2 U_{\phi\phi}\big] $.


\subsubsection*{Leptonic operators:}

After including the leptonic part in Eq.~\eqref{lep1}, we get Weinberg operators at dimension five and Four-Fermi operators at dimension six. The modulus square of the classical background field involving leptons can be written as
\begin{eqnarray}\label{eq:leptrpt}
    |\Delta_c|^2 &\supset& - \frac{1}{4 m_\Delta^4} (Y^{pq } _{\Delta})^\ast Y^{rs \,}_{\Delta} (\overline{l\,} ^{T \,p}_L C \, l^{\,s}_L)(\overline{l\,}^{T \,r}_L  C \, l^{\,q}_L)  -\frac{\mu_\Delta (Y^{pq}_{\Delta})^\ast}{2 m_\Delta^4} ( \tilde{H}^\dagger l_L^{\,p})(\overline{\tilde{l\,}}_L^{\,q} H ) + \text{ h.c} \, . 
\end{eqnarray}
Here, $p$ and $q$ represent the flavor indices and $C$ is the charge conjugation operator, $\tilde{l}_L = i \sigma_2 (l_L)^C =i \sigma_2 (l^C)_R= i \sigma_2 C \, (\overline{l\,}_L )^{T}$. Using the above Eq.~\eqref{eq:leptrpt},
the dimension six effective operators involving leptons, that we get at two-loop are 
\begin{align}
   & \dfrac{1}{2}\text{tr}\big[U_{\phi}^2\big] \;\quad \label{eq:trpferfirst}
     = {\dfrac{3}{4} \left(6\lambda^2_2 - \lambda^2_3 + 2 \lambda_2 \lambda_3\right) |\Delta_c|^2}
     \supset -\dfrac{3}{8 m_\Delta^4} \left(6\lambda^2_2 - \lambda^2_3 + 2 \lambda_2 \lambda_3\right) \n\\
     &\hspace{1.55cm}\times\Big( (Y^{pq } _{\Delta})^\ast Y^{rs \,}_{\Delta} (\overline{l} ^{T \,p}_L C \, l^{\,s}_L)(\overline{l}^{T \,r}_L  C \, l^{\,q}_L) \,+\,2 \mu_\Delta (Y^{pq}_{\Delta})^\ast( \tilde{H}^\dag l_L^{\,p})(\overline{\tilde{l}}_L^{\,\,q} H )
     + \text{ h.c} \Big) ,\\[0.4\baselineskip]
     &\dfrac{1}{2}\text{tr}\big[U U_{\phi\phi}\big]= \dfrac{1}{2} \left(2 \lambda_2 + \frac{3}{2} \lambda_3 \right) \big[(U_{11})_{ii} + (U_{22})_{ii}\big] 
      \supset -\frac{1}{4 m_\Delta^4} \left(2 \lambda_2 + \frac{3}{2} \lambda_3 \right)^2 \n\\
      &\hspace{1.9cm} \times\Big( (Y^{pq } _{\Delta})^\ast Y^{rs \,}_{\Delta} (\overline{l} ^{T \,p}_L C \, l^{\,s}_L)(\overline{l}^{T \,r}_L  C \, l^{\,q}_L) \,+\,2 \mu_\Delta (Y^{pq}_{\Delta})^\ast( \tilde{H}^\dag l_L^{\,p})(\overline{\tilde{l}}_L^{\,\,q} H )
     + \text{ h.c} \Big) .\label{eq:trpferend}
\end{align}
Finally, we present our results (only the two-loop part) in the following form,
\begin{eqnarray}\label{eq:twoloop}
    \mathcal{L} \supset \mathcal{O}_a \mathcal{C}_a,
\end{eqnarray}

where $\mathcal{O}_a$ denotes the dimension six pure bosonic and fermionic operators and $ \mathcal{C}_a$ denotes the corresponding Wilson coefficients. We use the model-independent result given in the Eqs.~\eqref{eq:effLag}-\eqref{eq:wilson-2} to write the effective action. Using that result, alsong with calculated results from the  Eqs.\eqref{eq:trpt1}-\eqref{eq:trpend}, we present the $\mathcal{C}_a$ corresponding to $\mathcal{O}_a$ in Tab.\,\ref{tbl:ct1} for bosonic case whereas using the Eqs.~\eqref{eq:trpferfirst}-\eqref{eq:trpferend}, we have listed the results in Tab.\,\ref{tbl:ct2} for leptonic case for the complex triplet model.

\begin{table}[ht]
\begin{center}
\renewcommand{\arraystretch}{1.65}
\begin{tabular}{|| c || c ||} 
\hline 
Dim six Ops. ($\mathcal{O}_a$)& Wilson coefficients ($\mathcal{C}_a$)  \\
\hline\hline
    $\mathcal{O}_6 $ & $ \frac{ 3\alpha^2  \mu_\Delta^2}{8 m_\Delta^4} \Big( \mathcal{C}^{[\![U U^2_{\phi }]\!]}_0 - 2\mathcal{C}^{[\![U^2_{\phi }]\!]}_2  \Big) \left( 6 \lambda_2^2 - \lambda_3^2 + 2 \lambda_2 \lambda_3 \right) \left(2\lambda_1+\lambda_4\right)   $ \\
    & $ - \frac{ 3\alpha^2  \mu_\Delta^2}{16 m_\Delta^4}  \mathcal{C}^{[\![U U_{\phi \phi}]\!]}_2 \left( 4 \lambda_2 + 3 \lambda_3\right)^2 \left(2\lambda_1+\lambda_4\right)   $ \\
    & $ + \frac{ \alpha^2\mu_\Delta^2}{4 m_\Delta^4} \mathcal{C}^{[\![ U^2 U_{\phi \phi }]\!]}_0 \left( 3 \lambda_2 + 2 \lambda_3 \right) \left(2\lambda_1\lambda_2+\frac{3\lambda_1\lambda_3}{2}+\lambda_2\lambda_4+\frac{3\lambda_3\lambda_4}{4}\right)  $\\
    & $ + \frac{\alpha^2}{16 m_\Delta^2} \mathcal{C}^{[\![ U^3 U_{\phi \phi }]\!]}_{-2} \left( 3 \lambda_2 + 2 \lambda_3 \right) \left(\lambda_1^3+\frac{3\lambda_4^3}{8}+\frac{3 \lambda_1^2\lambda_4}{2}+\frac{5\lambda_1\lambda_4^2}{4}\right) $\\[2pt]
    \hline
    $\mathcal{O}_H $ & $ \frac{3 \alpha^2\mu_\Delta^2}{2 m_\Delta^4} \Big( 2 \mathcal{C}^{[\![ U^2_{\phi }]\!]}_2 - \mathcal{C}^{[\![U_{\phi } D^2 U_{\phi }]\!]}_0  \Big) \left( 6 \lambda_2^2 - \lambda_3^2 + 2 \lambda_2 \lambda_3 \right)  $  \\ 
    & $ \frac{3\alpha^2\mu_\Delta^2}{2 m_\Delta^4}  \mathcal{C}^{[\![U U_{\phi \phi}]\!]}_2  \left( 4 \lambda_2 + 3 \lambda_3 \right)^2  $  \\ 
    & $ - \frac{\alpha^2}{m_\Delta^2} \mathcal{C}^{[\![U^2_{\mu } U_{\phi \phi}]\!]}_{-2} \left( 3 \lambda_2 + 2 \lambda_3 \right) \left(\frac{3\lambda_1^2}{4}+ \frac{3\lambda_4^2}{16}+\frac{3\lambda_1\lambda_4}{4}\right) $ \\[2pt]
    \hline
    $ \mathcal{O}_T $ & $ \frac{\alpha^2\lambda_4^2}{4 m_\Delta^2} \mathcal{C}^{[\![U^2_{\mu } U_{\phi \phi}]\!]}_{-2} \left( 3 \lambda_2 + 2 \lambda_3 \right) $ \\[2pt] 
    \hline
    $ \mathcal{O}_R $ & $ \frac{\alpha^2\lambda_4^2}{2 m_\Delta^2} \mathcal{C}^{[\![U^2_{\mu } U_{\phi \phi}]\!]}_{-2} \left( 3 \lambda_2 + 2 \lambda_3 \right) $ \\[2pt]
    \hline
    $ \mathcal{O}_{WW} $ & $ -\frac{\alpha^2}{4m_\Delta^2} \mathcal{C}^{[\![U G^2_{\mu \nu } U_{\phi \phi}]\!]}_{-2} \left( 4 \lambda_2 + 3 \lambda_3 \right) \left(2\lambda_1+\lambda_4\right) $ \\[2pt]
    \hline
    $ \mathcal{O}_{BB} $ & $ -\frac{3\alpha^2}{8m_\Delta^2} \mathcal{C}^{[\![U G^2_{\mu \nu } U_{\phi \phi}]\!]}_{-2} \left( 4 \lambda_2 + 3 \lambda_3 \right) \left(2\lambda_1+\lambda_4\right)$ \\[2pt]
    \hline
    $ \mathcal{O}_{WB} $ & $ \frac{\alpha^2\lambda_4}{2 m_\Delta^2} \mathcal{C}^{[\![U G^2_{\mu \nu } U_{\phi \phi}]\!]}_{-2} \left( 4 \lambda_2 + 3 \lambda_3 \right)$ \\[2pt]
    \hline
    $ \mathcal{O}_{2W} $ & $ -\frac{2\alpha^2 g^2_{W}}{m_\Delta^2} \mathcal{C}^{[\![J^2_{\mu } U_{\phi \phi}]\!]}_{-2} \left( 4 \lambda_2 + 3 \lambda_3 \right)$ \\[2pt]
    \hline
    $ \mathcal{O}_{2B} $ & $ -\frac{3 \alpha^2g^2_{Y}}{m_\Delta^2} \mathcal{C}^{[\![J^2_{\mu } U_{\phi \phi}]\!]}_{-2} \left( 4 \lambda_2 + 3 \lambda_3 \right)$ \\[2pt]
    \hline
    $ \mathcal{O}_{3W} $ & $ -\frac{3 \alpha^2g^2_{W}}{m_\Delta^2} \mathcal{C}^{[\![G^3 U_{\phi \phi}]\!]}_{-2} \left( 4 \lambda_2 + 3 \lambda_3 \right)$ \\[2pt]
\hline 
\end{tabular}
\caption{Dimension six CP-conserving pure bosonic operators and their corresponding two-loop Wilson coefficients for the complex triplet extension.}
\label{tbl:ct1}
\end{center}
\end{table}

\begin{table}[ht]
\begin{center}
\renewcommand{\arraystretch}{1.6}
\begin{tabular}{|| c ||  c ||} 
\hline 
Dim six Ops. ($\mathcal{O}_a$)& Wilson coefficients ($\mathcal{C}_a$)  \\
\hline 
$ (\overline{l\,} ^{T \,p}_L C \, l^{\,s}_L)(\overline{l\,}^{T \,r}_L  C \, l^{\,q}_L) + \text{h.c} $ & $ - \frac{3\alpha^2}{8 m_\Delta^2}(Y^{pq }_{\Delta})^\ast Y^{rs \,}_{\Delta}\mathcal{C}^{[\![U^2_{\phi }]\!]}_2 \left( 6 \lambda_2^2 - \lambda_3^2 + 2 \lambda_2 \lambda_3 \right) $ \\[2pt]
& $ - \frac{\alpha^2}{16 m_\Delta^2}(Y^{pq }_{\Delta})^\ast Y^{rs \,}_{\Delta} \mathcal{C}^{[\![U U_{\phi \phi}]\!]}_2 \left( 4 \lambda_2 + 3 \lambda_3 \right)^2 $ \\[2pt]
\hline\hline
\end{tabular}
\caption{Dimension six CP-conserving fermionic operators and their corresponding two-loop Wilson coefficients for complex triplet extension.}
\label{tbl:ct2}
\end{center}
\end{table}
%
%
\subsection{Electroweak doublet with hypercharge \texorpdfstring{$Y_\Phi=-\frac{1}{2}$}{Lg}}
Here, we concentrate on the scenario where the extra electroweak Higgs doublet $\Phi$ is integrated out to obtain two-loop effective action. The mass of the new scalar ($m_\Phi$) is assumed to be significantly higher than the electroweak symmetry-breaking scale \textit{i.e.,} $m_\Phi\gg v$. When $\Phi$ interacts with the SM Higgs doublet ($H$) and fermions, the most general UV Lagrangian can be expressed as~\cite{Henning:2014wua}
\begin{eqnarray}
    \mathcal{L}= \mathcal{L}_{\rm SM} + |D_\mu \Phi|^2- m_\Phi^2|\Phi|^2 - V(H,\Phi),
\end{eqnarray}
with potential term
\begin{eqnarray}\label{eq:Pot2HDM}
 V(H,\Phi) &=& \frac{\lambda_\Phi}{4}|\Phi|^4  -\big(\eta_H |\tilde{H}|^2 + \eta_\Phi |\Phi|^2\big)  \big(\tilde{H}^\dagger\Phi + \Phi^\dagger 
 \tilde{H}\big)
 +\lambda_1 |\tilde{H}|^2|\Phi|^2 + \lambda_2 |\tilde{H}^\dagger\Phi|^2 \\
 &+& \lambda_3 \big[\big(\tilde{H}^\dagger\Phi\big)^2 + \big(\Phi^\dagger\tilde{H}\big)^2 \big] 
    + \big( Y^{(e)}_{\Phi} \overline{l}_L \, \widetilde{\Phi} e_R + Y^{(u)}_{\Phi} \overline{q}_L \, \Phi \, u_R + Y^{(d)}_{\Phi} \overline{q}_L\, \widetilde{\Phi} \, d_R + \text{h.c.} \big)\n,
\end{eqnarray}
the covariant derivative $D_\mu$ has the same form as defined earlier for the complex triplet scalar. Here, Yukawa couplings, such as $ Y^{(e)}_{\Phi}, \,Y^{(u)}_{\Phi}$, and $Y^{(d)}_{\Phi} $, in general, can be complex, whereas other parameters are treated as real.\\
Analogous to the previous case, the two-loop contributions are computed relying on Eqs.~\eqref{eq:effLag}–\eqref{eq:wilson-2}. As discussed earlier, for this model also we need to calculate the $U$ matrix and its functional derivatives, $U_\phi$ and $U_{\phi\phi}$. The relevant details for this scenario is discussed as follows.

We can write the above potential in the following matrix form,
\begin{equation}
   \mathcal{L} \supset \dfrac{1}{2} \begin{pmatrix}
       \Phi^\ast_i & \Phi_i 
   \end{pmatrix} \, 
   U_{ij}
   \begin{pmatrix}
       \Phi_j \\ \Phi^\ast_j
   \end{pmatrix}
   = \dfrac{1}{2} \begin{pmatrix}
       \Phi^\ast_i & \Phi_i 
   \end{pmatrix} \, 
   \begin{pmatrix}
       \left(U_{11}\right)_{ij} & \left(U_{12}\right)_{ij} \\
       \left(U_{21}\right)_{ij} & \left(U_{22}\right)_{ij}
   \end{pmatrix} \,
   \begin{pmatrix}
       \Phi_j \\ \Phi^\ast_j
   \end{pmatrix},
\end{equation}
where $i,j$ runs from 1 to 2. The $4\times 4$ hermitian matrix $U$ is decomposed into four matrix elements, which are $2\times2$ matrices defined similarly as in the case of electroweak triplet model (see Eq. \eqref{eq:matrixtriplet}). The matrix elements are  

\begin{eqnarray}
        (U_{{11}})_{ij} & =& \dfrac{1}{2} \lambda_\Phi \Big[\, \big(\Phi^*_k \Phi_k \big) \delta_{ij} + \Phi_i \Phi^*_j \, \Big] -\eta_\Phi \Big[\, \big( \tilde{H}^*_k \Phi_k + \Phi^*_k \tilde{{H}_k} \big) \delta_{ij} + \Phi_i \tilde{H}^*_j + \tilde{{H}_i} \Phi^*_j \, \Big] \n \\
        & +& \lambda_1 \big(\, H^*_k H_k \, \big)\delta_{ij} + \lambda_2 \big(\, \tilde{H}_i \tilde{H}^*_j \, \big),
        \\[0.5\baselineskip]
        (U_{{12}})_{ij} & =&  \dfrac{1}{2} \lambda_{\Phi} \big( \Phi_i \Phi_j  \big) - \eta_\Phi \big(\Phi_i \tilde{H}_j + \tilde{H}_i \Phi_j \big) + \lambda_3 \Big[ \tilde{H}_i \tilde{H}_j + \tilde{H}_j \tilde{H}_i \Big], \\[0.5\baselineskip]
        (U_{{22}})_{ij} & =& (U_{{11}}^\ast)_{ij}, \quad (U_{{21}})_{ij}  = (U_{{12}}^\ast)_{ij}.
\end{eqnarray}
Using the CDE method, the classical background field (see Eq.~\eqref{eqn:phic_expand}) can be written as
\begin{eqnarray}
    \Phi_c^i = \frac{1}{m_\Phi^2}B^i + \frac{1}{m_\Phi^4} \left(p^2\delta_{ij}-\left(U_{11}\right)_{ij}\right) B^j +\mathcal{O}(m_\Phi^{-6}),
\end{eqnarray}
where $B^i = \eta_H | H |^2 \tilde{H}^i $ (see Eq.~\eqref{eqn:tree_UV_Lag}). Using this, we can write the following relations up to dimension six.
\begin{eqnarray}
    &|\Phi_c|^2 \label{eq:phisq1}
    & =(\Phi_c^i)^* \Phi_c^i \supset \dfrac{\eta^2_H}{m^4_\Phi} \mathcal{O}_6,\\
    &\Phi_c^\dagger\tilde{H}& + \tilde{H}^\dagger\Phi_c \supset \frac{2\eta_H}{m_\Phi^2}\left[|H|^4+\frac{1}{m_\Phi^2}\Big(\mathcal{O}_R+\mathcal{O}_H -\left(\lambda_1+\lambda_2\right)\mathcal{O}_6\Big)\right],\label{eq:phisq2}
\end{eqnarray}
where the dimension six operators are listed in Tab.\,\ref{tbl:operators}. We revisit the calculations for the one-loop correction at dimension six for this model given in App.~\ref{app:oneloop2HDM}, which can be verified with the results of \!~\cite{Henning:2014wua,Bakshi:2018ics}. 
\subsubsection{Operators contributing up to dimension six at two-loop level}
Here, we discuss the operators up to dimension six generated at the two-loop level for bosonic and fermionic cases for the doublet scenario. Before computing the operators, we would first define the 3-pt and 4-pt vertex that we defined earlier, for our model.
\begin{eqnarray}
 &&   (U_\phi)_{ijk} = {\frac{1}{2} \lambda_\Phi(\Phi_k \delta_{ij} + \Phi_i \delta_{jk}) -\eta_\Phi ( \tilde{H_k} \delta_{ij} + \tilde{H_i} \delta_{jk})},\n \\ 
&&    {(U^*_\phi)_{ijk}} = {\frac{1}{2} \lambda_\Phi(\Phi^*_k \delta_{ij} + \Phi^*_j \delta_{ik}) -\eta_\Phi ( \tilde{H^*_k} \delta_{ij} + \tilde{H^*_j} \delta_{ik})}, \\ \n
&&     {(U_{\phi \phi})_{ijkl}} = {\frac{1}{2} \lambda_\Phi(\delta_{ij} \delta_{kl} + \delta_{ik} \delta_{jl})}.
\end{eqnarray}

\subsubsection*{Bosonic operators:}
Using the expansion of the classical solution of the heavy doublet field, given in the Eqs.~\eqref{eq:phisq1}-\eqref{eq:phisq2}, the pure bosonic operators for this model are presented here.
\begin{eqnarray}
 &&  \frac{1}{2}\text{tr}\big[U_{\phi\phi}\big]
   =\label{eq:2hdm1st} 
   \frac{1}{2} \left[ \frac{\partial^2 (U_{11} )_{ii}}{\partial\Phi_k\partial\Phi_k^\dagger} +  \frac{\partial^2 (U_{22} )_{ii}}{\partial\Phi_k\partial\Phi_k^\dagger} \right] = 3 \lambda_\Phi , \\
   [0.4\baselineskip]
&&   \frac{1}{2}\text{tr}\big[U_{\phi}^2\big] \label{eq:2hdm2}
   =
   3 \left[\frac{\partial (U_{11})_{ij} }{\partial\Phi_m}\frac{\partial (U_{22})_{ji} }{\partial\Phi_m^\dagger} \right] , \n\\
   &&\hspace{15mm}=  \frac{3}{4}\left[5 \lambda_\Phi^2 |\Phi_c|^2 - 10 \lambda_\Phi\eta_\Phi \left(\Phi_c^\dagger\tilde{H} + \tilde{H}^\dagger\Phi_c\right) + 24 \eta _{\Phi }^2 | H |^2 \right], \n \\
   &&\hspace{15mm}\supset \frac{3}{4}\Big[\dfrac{5\eta^2_H \lambda_\Phi^2}{m^4_\Phi}  \mathcal{O}_6 - \frac{20\eta_H}{m_\Phi^2}\lambda_\Phi\eta_\Phi\left(|H|^4+\frac{1}{m_\Phi^2}\Big(\mathcal{O}_R+\mathcal{O}_H -\left(\lambda_1+\lambda_2\right)\mathcal{O}_6\Big)\right) \n\\
   &&\hspace{15mm}+ 24 \eta _{\Phi }^2 | H |^2 \Big] , \\
   [0.4\baselineskip]
&&   \frac{1}{2}\text{tr}\big[U U_{\phi\phi}\big] \label{eq:2hdm3}
   =
   \frac{3\lambda_\Phi}{4}\left[\big(U_{11}\big)_{ii} +\big(U_{22}\big)_{ii} \right], \n \\
   &&\hspace{18mm}=  \frac{3\lambda _{\Phi }}{4}\Big[\left(4  \lambda _1+2  \lambda _2\right)|H|^2-6 \eta _{\Phi } \left(\Phi_c^\dagger\tilde{H} + \tilde{H}^\dagger\Phi_c\right)+3  \lambda _{\Phi }|\Phi_c|^2 \Big]\n\\ 
   &&\hspace{18mm}\supset  \frac{3\lambda _{\Phi }}{4} \Big[\left(4 \lambda _1 + 2 \lambda _2\right)|H|^2 - \dfrac{12}{m^2_\Phi}  \eta_\Phi \eta_H  |H|^4 \n\\
   &&\hspace{18mm}+ \dfrac{3}{ m^4_\Phi } \Big( \lambda_\Phi \eta^2_H + 4 \eta_\Phi \eta_H (\lambda_1 + \lambda_2) \Big)\mathcal{O}_6 - \dfrac{12}{m_\Phi^4} \eta_\Phi \eta_H  (\mathcal{O}_H + \mathcal{O}_R) \Big] ,\\
   [0.4\baselineskip]
 &&  \dfrac{1}{2}\text{tr}\big[UU_{\phi}^2\big] \label{eq:2hdm4}
   =  \dfrac{1}{2} \bigg[ \frac{\partial \, (U_{11})_{ij}}{\partial \Phi_k} (U_{11})_{ii'} \frac{\partial \,  (U_{11})_{i'j}}{\partial \Phi_k^\ast} + \frac{\partial \,  (U_{22})_{ij} }{\partial \Phi_k^\ast} (U_{22})_{ii'} \frac{\partial \,  (U_{22})_{i'j} }{\partial \Phi_k} \n \\[0.15\baselineskip]
      &&\hspace{18mm}+  \frac{\partial \,  (U_{11})_{ij} }{\partial \Phi_k} (U_{12})_{ii'} \frac{\partial \, (U_{11})_{i'j} }{\partial \Phi_k} + \frac{\partial \, (U_{22})_{ij} }{\partial \Phi_k^\ast} (U_{21})_{ii'} \frac{\partial \, (U_{22})_{i'j} }{\partial \Phi_k^\ast} \bigg], \n \\[0.15\baselineskip]
   &&\hspace{18mm}=  \frac{3}{2}\Big[ \eta _{\Phi }^2 \left(6 \lambda _1 + 4 \lambda _2 + 4 \lambda_3 \right) |H|^4-  9 \eta_{\Phi }^3 \left(\Phi_c^\dagger\tilde{H} + \tilde{H}^\dagger\Phi_c\right)|H|^2 \n \\
   &&\hspace{18mm}-  \eta _{\Phi }\lambda _{\Phi }\big( 5 \lambda _1 +  4 \lambda _2 + 4 \lambda_3\big) \left(\Phi_c^\dagger\tilde{H} + \tilde{H}^\dagger\Phi_c\right)|H|^2 \Big] \n\\
   &&\hspace{18mm}\supset \frac{3}{2}\Big[\eta_{\Phi }^2 \Big( 6 \lambda _1 + 4 \lambda _2 + 4 \lambda_3 \Big) |H|^4 \n \\
   &&\hspace{18mm}- \frac{2\eta_H}{m_\Phi^2}\Big(10\eta_\Phi^3 + \eta _{\Phi }\lambda _{\Phi }\big( 5 \lambda _1 + 4 \lambda _2 + 4 \lambda_3\big)\Big)  \mathcal{O}_6  \Big] , \\
   [0.4\baselineskip]
&&  \frac{1}{2}\text{tr}\big[U^2 U_{\phi\phi}\big] \label{eq:2hdm5}
  =  \frac{3\lambda_\Phi}{4}\left[\big(U_{11}^2\big)_{ii} +2\big(U_{12}\big)_{ij} \big(U_{21}\big)_{ji} +\big(U^2_{22}\big)_{ii} \right], \n \\
   &&\hspace{21mm}=  \frac{3\lambda _{\Phi }}{4}\Big[ \big(4 \lambda _1^2+ 4 \lambda _1 \lambda _2+ 2 \lambda _2^2+ 8 \lambda _3^2\big)|H|^4 \n\\
   &&\hspace{21mm}- 4 \eta _{\Phi } \big(3 \lambda _1+2 \left(\lambda _2+ \lambda _3\right)\big)  \left(\Phi_c^\dagger\tilde{H} + \tilde{H}^\dagger\Phi_c\right)|H|^2\Big] \n \\
   &&\hspace{21mm}\supset  \frac{3\lambda _{\Phi }}{4}\Big[ \Big(4 \lambda _1^2+4 \lambda _2 \lambda _1+ 2\lambda _2^2+ 8\lambda _3^2\Big)|H|^4 \n\\
   &&\hspace{21mm}- \frac{8\eta_\Phi\eta_H }{m_\Phi^2}  \Big(3 \lambda _1+2 \big(\lambda _2+\lambda _3\big)\Big)  \mathcal{O}_6\Big] , \\
   \n \\[0.4\baselineskip]
&&    \frac{1}{2}\text{tr}\big[U_{\phi}D^2U_{\phi}\big] =   \dfrac{1}{2} \left[\frac{\partial \,  U_{ij}}{\partial \Phi_k}D^2\frac{\partial \,  U_{ji}}{\partial \Phi_k^\ast}\right]= -\dfrac{1}{2} \left[D_\mu\left(\frac{\partial \, U_{ij}}{\partial \Phi_k}\right)D^\mu\left(\frac{\partial \, U_{ji}}{\partial \Phi_k^\ast}\right)\right], \n \\
    &&\hspace{25mm}\supset -\frac{3}{4}\Big[20 \eta_\Phi^2 |D_\mu H|^2 - \dfrac{20}{m^2_\Phi} \lambda_\Phi \eta_\Phi \eta_H \big( \mathcal{O}_H + \mathcal{O}_R \big)\Big] , \\
   [0.4\baselineskip]
&&  \frac{1}{2} \text{tr}\big[G_{\mu\nu}^2U_{\phi\phi}\big]
  = \frac{3\lambda_\Phi}{4}\text{tr}[G_{\mu\nu}^2]  
   = -\dfrac{3\lambda_\Phi}{4}  \left( \frac{1}{2}g_W^2 \left(W_{\mu\nu}^a\right)^2 + \frac{1}{2}g_Y^2 \left(B_{\mu\nu}\right)^2 \right) , \\
   [0.4\baselineskip]
&&  \dfrac{1}{2}\text{tr}\big[U^2U_{\phi}^2\big]
 =  \dfrac{1}{2} \bigg[ \frac{\partial \, (U_{11})_{ij} }{\partial \Phi_k} \big[U^2_{11}+U_{12}U_{21}\big]_{ii'} \frac{\partial \, (U_{11})_{i'j}}{\partial \Phi_k^\ast} +  \frac{\partial \, (U_{22})_{ij} }{\partial \Phi_k} \big[U_{21}(U_{11}+U_{22})\big]_{ij} \frac{\partial \, (U_{22})_{i'j}}{\partial \Phi_k^\ast} \n \\[0.15\baselineskip]
    &&\hspace{15mm}+ \frac{\partial \, (U_{11})_{ij} }{\partial \Phi_k} \big[U_{12}(U_{11}+U_{22})\big]_{ij} \frac{\partial \, (U_{11})_{i'j}}{\partial \Phi_k^\ast} +  \frac{\partial \, (U_{22})_{ij} }{\partial \Phi_k} \big[U^2_{22}+U_{21}U_{12}\big]_{ij} \frac{\partial \, (U_{22})_{i'j}}{\partial \Phi_k^\ast} \bigg] \n\\
   &&\hspace{15mm}\supset \frac{3}{2}\left[\eta _{\Phi }^2   \Big( 6\lambda_1 + 4 \lambda_2 + 4 \lambda_3\Big)^2\right] \mathcal{O}_6 , \\
   [0.4\baselineskip]
 && \frac{1}{2} \text{tr}\big[U^3U_{\phi\phi}\big] =  \frac{3\lambda_\Phi}{4} \big[\big(U_{11}^3+3 U_{11}U_{12}U_{21}+3 U_{22}U_{21}U_{12}+U^3_{22}\big)\big]_{ii} \n \\
   &&\hspace{22mm}\supset \frac{3\lambda_\Phi}{4} \Big[4 \lambda _1^3+6 \lambda _2 \lambda _1^2+6 \lambda _2^2 \lambda _1 +2 \lambda _2^3 +24\lambda _1 \lambda _3^2 +24 \lambda _2 \lambda _3^2 \Big] \mathcal{O}_6 , \\
   [0.4\baselineskip]
&&  \frac{1}{2}\text{tr}\big[ G^2_{\mu \nu} U^2_{\phi} \big] =  \frac{1}{2}\left[ \frac{\partial (U_{ij})}{\partial\Phi_k} \! \left(G^2_{\mu \nu}\right)_{ii'} \! \frac{\partial (U_{i'j})}{\partial\Phi_k^\dagger} \right]  
  \! \supset - \frac{1}{2}\Big[6\eta_\Phi^2 \Big( \mathcal{O}_{WW} + \mathcal{O}_{BB}  + 2\mathcal{O}_{WB}\Big) \! \Big], \\
  [0.3\baselineskip]
&&  \frac{1}{2} \text{tr} \big[U^2_{\phi} U_{\mu \mu} \big] =  \frac{1}{2} \left[ \frac{\partial (U_{ij})}{\partial\Phi_k} \left(D^2 U \right)_{ii'} \frac{\partial (U_{i'j})}{\partial\Phi_k^\dagger} \right] \supset - \frac{3}{2} \eta _{\Phi }^2 \left(6\lambda _1+ 4\lambda _2+ 4\lambda_3 \right) \mathcal{O}_H \\[0.3\baselineskip]
 && \frac{1}{2}\text{tr} \big[ U G_{\mu \nu}^2 U_{\phi \phi} \big] = \frac{3\lambda_\Phi}{4} \text{tr} \big[ U G_{\mu \nu}^2 \big] 
  \! \supset - \frac{3\lambda_\Phi}{4} \Big[ \Big(\lambda_1 + \dfrac{\lambda_2}{2}\Big)\Big(\mathcal{O}_{WW} + \mathcal{O}_{BB}\Big) +  \lambda_2 \mathcal{O}_{WB}\Big] , \\
    [0.4\baselineskip]
&&  \frac{1}{2} \text{tr} \big[ G_{\mu\nu}^3 U_{\phi \phi} \big] =\frac{3\lambda_\Phi}{4} \text{tr} \big[ G_{\mu\nu}^3 \big] 
  \! = - \dfrac{3\lambda_\Phi}{4} \Big( 3 g_W^2 \mathcal{O}_{3W}\Big) , \\
  [0.4\baselineskip]
&& \frac{1}{2} \text{tr} \big[ J_{\nu}^2 U_{\phi \phi} \big] = \frac{3\lambda_\Phi}{4} \text{tr} \Big[(D_\mu G_{\mu \nu})^2 \Big]  
  \! = - \frac{3\lambda_\Phi}{4} \Big[2g^2_W \mathcal{O}_{2W} + 2 g^2_Y \mathcal{O}_{2B}\Big] , \\
  [0.4\baselineskip]
 && \frac{1}{2} \text{tr} \big[U_\mu^2 U_{\phi \phi} \big] =\frac{3\lambda_\Phi}{4} \big[(D_\mu U)^2 \big]  
  \! \supset \frac{3\lambda_\Phi}{4} \Big[2 \Big(4\lambda_1^2 + \lambda_2^2 + 4\lambda_3^2 + 4\lambda_1 \lambda_2\Big) \mathcal{O}_H \n \\
  &&\hspace{20mm}+ 2\Big(\lambda_2^2 -4 \lambda_3^2\Big) \mathcal{O}_T + 4\Big(\lambda_2^2 + 4 \lambda_3^2\Big)\mathcal{O}_R \Big] , \\
  [0.4\baselineskip]
&&  \frac{1}{2}\text{tr}\big[U U_{\mu\mu} U_{\phi \phi} \big] = -\frac{3\lambda_\Phi}{4} \big[(D_\mu U)^2 \big]  
  \! \supset -\frac{3\lambda_\Phi}{4} \Big[2 \Big(4\lambda_1^2 + \lambda_2^2 + 4\lambda_3^2 + 4\lambda_1 \lambda_2\Big) \mathcal{O}_H \n \\
  &&\hspace{25mm}+ 2\Big(\lambda_2^2 -4 \lambda_3^2\Big) \mathcal{O}_T + 4\Big(\lambda_2^2 + 4 \lambda_3^2\Big)\mathcal{O}_R \Big],\label{eq:2hdmend}
\end{eqnarray}

We used the relations given in Eqs.~\eqref{app:essential1}-\eqref{app:essential2} to simplify and present our result in the SILH basis operators listed in Tab.\,\ref{tbl:operators}.

Note that in a doublet scenario, the operators $U^2 U_\phi^2$, $U_{\mu \mu} U_\phi^2$, and $G_{\mu \nu}^2 U_\phi^2$ do have contributions at dimension six, unlike in the case of complex triplet. 
\subsubsection*{Fermionic operators:}
Now, we turn to the fermionic part of the Lagrangian mentioned above (see Eq.~\eqref{eq:Pot2HDM}) for this model. For the classical background field $\Phi_c$ involving leptons and quarks, we can write the following relations
\begin{eqnarray}\label{eq:lep2hdm}
  &&   |\Phi_c|^2  \supset - \dfrac{1}{m_\Phi^4} \Big( Y^{(u)}_{\Phi} Y^{(d)}_{\Phi} (\overline{q}^{\,j}_{L} u_R) \epsilon^{jk} (\overline{q}^{\,k}_{L} d_R)  - \eta_H Y^{(e)}_\Phi (\overline{l}_L \tilde{H}^\dag e_R)|H|^2 \n \\
     &&\hspace{10mm}- \eta_H Y^{(u)}_\Phi (\overline{q}_L H u_R) |H|^2 - \eta_H Y^{(d)}_\Phi (\overline{q}_L \tilde{H}^\dag d_R)|H|^2  +  \text{h.c} \Big),\\[0.4\baselineskip]
&&     \Phi_c^\dagger\tilde{H} + \tilde{H}^\dagger\Phi_c \supset\frac{\eta_H}{m_\Phi^2} \Big( Y^{(e)}_\Phi (\overline{l}_L \tilde{H}^\dag e_R) 
     +  Y^{(u)}_\Phi (\overline{q}_L H u_R) 
      +  Y^{(d)}_\Phi (\overline{q}_L \tilde{H}^\dag d_R)  +  \text{h.c} \Big),
\end{eqnarray}
where $\epsilon^{ij} = i (\sigma_2)^{ij}$. Using the above two equations, and Eqs.~\eqref{eq:2hdm2}-\eqref{eq:2hdm5} in the effective Lagrangian given in \eqref{eq:effLag}, the operators involving fermions, that we get at two-loop are

\begin{align}
    &\frac{1}{2}\text{tr}\big[U_{\phi}^2\big]\label{eq:2hdmfer1st}  = \frac{1}{4}\left(15 \lambda_\Phi^2 |\Phi_c|^2 
   \right)
   \supset - \frac{15 \lambda_\Phi^2}{4 m_\Phi^4} \Big( Y^{(u)}_{\Phi} Y^{(d)}_{\Phi} (\overline{q}^{\,j}_{L} u_R) \epsilon^{jk} (\overline{q}^{\,k}_{L} d_R)  - \eta_H Y^{(e)}_\Phi (\overline{l}_L \tilde{H}^\dag e_R)|H|^2 \n \\
     & \hspace{1.8cm}- \eta_H Y^{(u)}_\Phi (\overline{q}_L H u_R) |H|^2 - \eta_H Y^{(d)}_\Phi (\overline{q}_L \tilde{H}^\dag d_R)|H|^2  +  \text{h.c} \Big), \\
    & \frac{1}{2}\text{tr}\big[U U_{\phi\phi}\big]
   \,\, =  \frac{9\lambda _{\Phi }^2}{4} 
    |\Phi_c|^2 
   \supset -\frac{9\lambda _{\Phi }^2}{4 m_\Phi^4} \Big( Y^{(u)}_{\Phi} Y^{(d)}_{\Phi} (\overline{q}^{\,j}_{L} u_R) \epsilon^{jk} (\overline{q}^{\,k}_{L} d_R)  - \eta_H Y^{(e)}_\Phi (\overline{l}_L \tilde{H}^\dag e_R)|H|^2 \n \\
     & \hspace{1.99cm}- \eta_H Y^{(u)}_\Phi (\overline{q}_L H u_R) |H|^2 - \eta_H Y^{(d)}_\Phi (\overline{q}_L \tilde{H}^\dag d_R)|H|^2  +  \text{h.c} \Big) ,\\
    & \frac{1}{2}\text{tr}\big[U^2 U_{\phi\phi}\big] = -3 \lambda_{\Phi} \eta_{\Phi} \Big[\big(3 \lambda _1+2 \left(\lambda _2+ \lambda _3\right)\big)  \left(\Phi_c^\dagger\tilde{H} + \tilde{H}^\dagger\Phi_c\right)|H|^2\Big]\n \\
     & \hspace{2.0cm} \supset -3 \dfrac{\lambda_{\Phi} \eta_{\Phi}\eta_H}{m_\Phi^2} \big(3 \lambda _1+2 \left(\lambda _2+ \lambda _3\right)\big) \Big( Y^{(e)}_\Phi (\overline{l}_L \tilde{H}^\dag e_R)|H|^2 
     +  Y^{(u)}_\Phi (\overline{q}_L H u_R) |H|^2 \n \\ 
     & \hspace{2.0cm}+ Y^{(d)}_\Phi (\overline{q}_L \tilde{H}^\dag d_R)|H|^2  +  \text{h.c} \Big) .\label{eq:2hdmferend}
\end{align}
Using the Eqs.~\eqref{eq:wilson+4}-\eqref{eq:wilson-2} and Eqs.\eqref{eq:2hdm1st}-\eqref{eq:2hdmend}, we present the $\mathcal{C}_a$ corresponding to $\mathcal{O}_a$ (see Eq.~\eqref{eq:twoloop}) in Tab.~\ref{tbl:2hdm1} for bosonic case, whereas using the Eqs.~\eqref{eq:wilson+4}-\eqref{eq:wilson-2} and Eqs.~\eqref{eq:2hdmfer1st}-\eqref{eq:2hdmferend}, we have listed the results in Tab.~\ref{tbl:2hdm2} for fermionic case for this scenario.

\begin{table}[ht]
\begin{center}
\renewcommand{\arraystretch}{1.65}
\begin{tabular}{|| c || c ||} 
\hline 
Dim six Ops. ($\mathcal{O}_a$)& Wilson coefficients ($\mathcal{C}_a$)  \\
\hline\hline
    $\mathcal{O}_6 $ \! \! & $  \bigg( 15 \mathcal{C}^{[\![U^2_{\phi}]\!]}_{2} + 9 \mathcal{C}^{[\![U U_{\phi \phi}]\!]}_{2} \bigg) \Big(\frac{ \alpha^2\, \eta^2_H \lambda_\Phi^2}{4 m^2_\Phi} - \frac{\alpha^2 \, \eta_H \lambda_\Phi\eta_\Phi}{m_\Phi^2} \big( \lambda_1 + \lambda_2 \big)\Big) $\\
    & $ - \mathcal{C}^{[\![U U^2_{\phi}]\!]}_{0} \frac{3\alpha^2\eta_H \eta_\Phi}{m_\Phi^2}\Big(10\eta_\Phi^2 + \lambda _{\Phi }\big( 5 \lambda _1 + 4\lambda _2 + 4\lambda_3\big)\Big)$ \\
    & $ - \mathcal{C}^{[\![U^2 U_{\phi \phi}]\!]}_{0} \frac{6 \alpha^2\lambda_\phi \eta_H \eta_\Phi}{m_\Phi^2}\Big(3 \lambda_1 + 2 \lambda _2 + 2\lambda_3\Big) - \mathcal{C}^{[\![U^2 U^2_{\phi}]\!]}_{-2} \frac{6\alpha^2 \eta^2_\Phi}{m_\Phi^2}\Big(3\lambda_1 + 2 \lambda _2 + 2 \lambda_3\Big)^2$ \\
    & $ + \, \mathcal{C}^{[\![U^3 U_{\phi \phi}]\!]}_{-2} \frac{3\alpha^2\lambda_\Phi}{4 m^2_\Phi} \Big(4 \lambda _1^3+6 \lambda _2 \lambda _1^2+6 \lambda _2^2 \lambda _1 +2 \lambda _2^3 +24\lambda _1 \lambda _3^2 +24 \lambda _2 \lambda _3^2 \Big)$ \\[2pt]
    \hline
    $\mathcal{O}_H $  \! \! & $  \bigg(15 \mathcal{C}^{[\![U_{\phi} D^2 U_{\phi}]\!]}_{0}  - 15 \mathcal{C}^{[\![U^2_{\phi}]\!]}_{2} - 9\mathcal{C}^{[\![U U_{\phi \phi}]\!]}_{2} \bigg) \frac{ \alpha^2\, \eta_H \lambda_\Phi\eta_\Phi}{m_\Phi^2} $  \\
    & $- \mathcal{C}^{[\![U_{\mu\mu}U^2_{\phi}]\!]}_{-2} \frac{3\alpha^2 \eta _{\Phi }^2 }{m_\Phi^2}\Big(3\lambda _1+ 2\lambda _2+2\lambda_3 \Big) + \mathcal{C}^{[\![U^2_{\mu} U_{\phi \phi}]\!]}_{-2} \frac{3\alpha^2\lambda_\Phi}{ m^2_\Phi} \Big(4\lambda_1^2 + \lambda_2^2 + 4\lambda_3^2 + 4\lambda_1 \lambda_2\Big)$ \\[2pt]
    \hline
    $ \mathcal{O}_T $  \! \! & $ \mathcal{C}^{[\![U^2_{\mu} U_{\phi \phi}]\!]}_{-2} \frac{3\alpha^2\lambda_\Phi}{ m^2_\Phi} \Big(\lambda_2^2 - 4\lambda_3^2 \Big) $ \\[2pt]
    \hline
    $ \mathcal{O}_R $  \! \! & $  \bigg(15 \mathcal{C}^{[\![U_{\phi} D^2 U_{\phi}]\!]}_{0}  - 15 \mathcal{C}^{[\![U^2_{\phi}]\!]}_{2} - 9\mathcal{C}^{[\![U U_{\phi \phi}]\!]}_{2} \bigg) \frac{\alpha^2 \, \eta_H \lambda_\Phi\eta_\Phi}{m_\Phi^2} $ \\
    & $+\mathcal{C}^{[\![U^2_{\mu} U_{\phi \phi}]\!]}_{-2}  \frac{6\alpha^2\lambda_\Phi}{m^2_\Phi} \Big( \lambda_2^2 + 4\lambda_3^2 \Big)$ \\[2pt]
    \hline
    $ \mathcal{O}_{WW} $  \! \! & $ - \mathcal{C}^{[\![G^2_{\mu \nu} U^2_{\phi}]\!]}_{-2} \frac{9\alpha^2\eta_\Phi^2}{2 m^2_\Phi} - \mathcal{C}^{[\![U G^2_{\mu \nu} U_{\phi \phi}]\!]}_{-2} \frac{3\alpha^2\lambda_\Phi}{8 m^2_\Phi} \Big(2\lambda_1 + \lambda_2\Big) $ \\[2pt]
    \hline
    $ \mathcal{O}_{BB} $  \! \! & $ - \mathcal{C}^{[\![G^2_{\mu \nu} U^2_{\phi}]\!]}_{-2} \frac{9\alpha^2\eta_\Phi^2}{2 m^2_\Phi} - \mathcal{C}^{[\![U G^2_{\mu \nu} U_{\phi \phi}]\!]}_{-2} \frac{3\alpha^2\lambda_\Phi}{8 m^2_\Phi} \Big(2\lambda_1 + \lambda_2\Big) $ \\[2pt]
    \hline
    $ \mathcal{O}_{WB} $  \! \! & $ - \mathcal{C}^{[\![G^2_{\mu \nu} U^2_{\phi}]\!]}_{-2} \frac{9\alpha^2\eta_\Phi^2}{ m^2_\Phi} - \mathcal{C}^{[\![U G^2_{\mu \nu} U_{\phi \phi}]\!]}_{-2} \frac{3\alpha^2\lambda_\Phi}{4 m^2_\Phi} \Big(2\lambda_1 + \lambda_2\Big) $ \\[2pt]
    \hline
    $ \mathcal{O}_{2W} $  \! \! & $ - \mathcal{C}^{[\![J^2_{\mu} U_{\phi \phi}]\!]}_{-2} \frac{3\alpha^2\lambda_\Phi}{2 m^2_\Phi} g^2_W $ \\[2pt]
    \hline
    $ \mathcal{O}_{2B} $  \! \! & $ - \mathcal{C}^{[\![J^2_{\mu} U_{\phi \phi}]\!]}_{-2} \frac{3\alpha^2\lambda_\Phi}{2 m^2_\Phi} g^2_Y$ \\[2pt]
    \hline
    $ \mathcal{O}_{3W} $  \! \! & $ - \mathcal{C}^{[\![G^3 U_{\phi \phi}]\!]}_{-2} \frac{9\alpha^2\lambda_\Phi}{4 m^2_\Phi} g^2_W$ \\[2pt]
\hline 
\end{tabular}
\caption{Dimension six CP-conserving pure bosonic operators and their corresponding two-loop Wilson coefficients for extra Higgs doublet extension.}
\label{tbl:2hdm1}
\end{center}
\end{table}

\begin{table}[ht]
\begin{center}
\renewcommand{\arraystretch}{1.65}
\begin{tabular}{|| c ||  c ||} 
\hline 
Dim six Ops. ($\mathcal{O}_a$) & Wilson coefficients ($\mathcal{C}_a$)  \\
\hline\hline
$ (\overline{q}^{\,j}_{L} u_R) \epsilon^{jk} (\overline{q}^{\,k}_{L} d_R) + \text{ h.c}  $ & $ - \frac{\alpha^2 \lambda_\Phi^2}{4 m_\Phi^2} Y^{(u)}_{\Phi} Y^{(d)}_{\Phi} \Big( 9 \mathcal{C}^{[\![U U_{\phi \phi}]\!]}_2 + 15\mathcal{C}^{[\![U^2_{\phi }]\!]}_2  \Big) $ \\[2pt]
\hline
$ (\overline{l}_L \tilde{H}^\dag e_R)|H|^2 + \text{h.c}$ & $ \frac{\alpha^2 \lambda_\Phi^2}{4 m_\Phi^2} Y^{(e)}_{\Phi} \Big(9 \mathcal{C}^{[\![U U_{\phi \phi}]\!]}_2 + 15\mathcal{C}^{[\![U^2_{\phi }]\!]}_2  \Big) $\\
& $ - \frac{3 \alpha^2\lambda_\Phi \eta_H \eta_\Phi}{m_\Phi^2} Y^{(e)}_\Phi  \mathcal{C}^{[\![U^2 U_{\phi \phi}]\!]}_{0} \Big(3 \lambda_1 + 2\big( \lambda _2 + \lambda_3\big)\Big)$\\[2pt]
\hline
$ (\overline{q}_L H u_R) |H|^2 +  \text{h.c}$ & $ \frac{ \alpha^2\lambda_\Phi^2}{4 m_\Phi^2} Y^{(u)}_{\Phi}  \Big(9 \mathcal{C}^{[\![U U_{\phi \phi}]\!]}_2 + 15\mathcal{C}^{[\![U^2_{\phi }]\!]}_2  \Big) $\\
& $ - \frac{3 \alpha^2\lambda_\Phi \eta_H \eta_\Phi}{m_\Phi^2} Y^{(u)}_\Phi \mathcal{C}^{[\![U^2 U_{\phi \phi}]\!]}_{0} \Big(3 \lambda_1 + 2\big( \lambda _2 + \lambda_3\big)\Big)$\\[2pt]
\hline
$ (\overline{q}_L \tilde{H}^\dag d_R)|H|^2  + \text{h.c}$ & $ \frac{ \alpha^2\lambda_\Phi^2}{4 m_\Phi^2} Y^{(d)}_{\Phi} \Big(9 \mathcal{C}^{[\![U U_{\phi \phi}]\!]}_2 + 15\mathcal{C}^{[\![U^2_{\phi }]\!]}_2  \Big) $\\
& $ - \frac{3\alpha^2 \lambda_\Phi \eta_H \eta_\Phi}{m_\Phi^2} Y^{(d)}_\Phi \mathcal{C}^{[\![U^2 U_{\phi \phi}]\!]}_{0} \Big(3 \lambda_1 + 2\big( \lambda _2 + \lambda_3\big)\Big)$\\[2pt]
\hline 
\end{tabular}
\caption{Dimension six CP-conserving fermionic operators and their corresponding two-loop Wilson coefficients for extra Higgs doublet extension.}
\label{tbl:2hdm2}
\end{center}
\end{table}
\section{Conclusions} \label{con} The goal of EFT is to systematically understand the low-energy behaviour of a UV theory through a set of parameters that can be measured in the experiments. The top-down approach involves integrating out the heavy particles with masses above the energy scale of interest, e.g., the electroweak symmetry-breaking scale. The effect of these heavy particles is captured in the Wilson coefficients corresponding to higher-dimensional operators. In this precision era of current and future collider experiments, it is necessary to go beyond one-loop corrections. Thus, it is important to compute the effective action in two-loop order, and that also signifies that EFT calculations are more closely mimicking the full theory computation effectively.

In this paper, we have applied the Heat-Kernel (HK) method to calculate the two-loop effective action. Using HKCs, we have defined an interacting Green's function that is free from divergences at the coincidence limit. We have computed the distinct irreducible vacuum diagrams, consisting of the interacting Green's functions and the vertex factors. We have validated the divergent part of each of the vacuum diagrams with the Ref.~\cite{Banerjee:2024rbc}. In the process, we have been able to extract the finite parts, to compute the two-loop effective action upto dimension six,  which is the primary aim of this paper.
We first consider a quantum field theory for scalars with a general interaction, $U$, and compute the two-loop corrections, involving only the heavy degrees of freedom for the individual diagrams. Then, based on our generic prescription, we have calculated the two-loop effective action up to dimension six for two example scenarios: when the SM is extended by an electroweak triplet $\Delta$ with hypercharge $Y_\Delta=1$ and the extension of the SM by an extra Higgs doublet $\Phi$ with hypercharge $Y_\Phi=-1/2$. For these two cases, we have computed the Wilson coefficients, which are functions of the parameters of the UV Lagrangian, corresponding to the dimension six pure bosonic as well as fermionic operators. For the sake of completeness, we have also noted the corrections to the lower-dimensional operators, e.g.,  $|H|^2,\, |D_{\mu}H|^2, \,|H|^4, \,(W_{\mu\nu}^a)^2,\, (B_{\mu\nu})^2$. 
Our paper presents the following results, highlighting the main findings of our work.
\begin{itemize}[topsep=-0.5ex,itemsep=-0.5ex,partopsep=0.5ex,parsep=0.5ex]
    \item We have computed the two-loop effective action in a model-independent manner by integrating out a heavy scalar, considering only the loops involving the heavy particle.
    \item We have calculated the dimension six SMEFT operators generated at two-loop, for the two cases, which are: 
    \begin{enumerate}[topsep=-1.25ex,itemsep=-0.75ex,partopsep=1ex,parsep=1ex]
           \item SM + electroweak scalar triplet with hypercharge $Y_\Delta=1$:
           \begin{enumerate}
               \item Dimension six bosonic operators with their corresponding Wilson coefficients are listed in Tab.~\ref{tbl:ct1}
               \item Dimension six leptonic operators with their corresponding Wilson coefficients are listed in Tab.~\ref{tbl:ct2}
           \end{enumerate}
           \item SM + electroweak scalar doublet with hypercharge $Y_\Phi = -1/2$:
           \begin{enumerate}
               \item Dimension six bosonic operators with their corresponding Wilson coefficients are listed in Tab.~\ref{tbl:2hdm1}
               \item Dimension six fermionic operators with their corresponding Wilson coefficients are listed in Tab.~\ref{tbl:2hdm2}
           \end{enumerate}
      \end{enumerate}
\end{itemize}

\paragraph{}
We have omitted heavy-light mixing contributions in this work, leaving their analysis at the two-loop level for future studies.

Though we have discussed the two-loop computation for integrating out heavy scalars, following the footsteps of Ref.~ \cite{Chakrabortty:2023yke}, we can suitably extend this result, i.e., the algebraic form for the effective Lagrangian to compute the same in the case of heavy fermion integrating out. We leave that part for future work, which will be more relevant regarding the emergence of CP-violating effective operators; see Ref.~\cite{Bakshi:2021}.   Moreover, this model-independent approach is systematic enough to be automated and streamlined. Thus, the computation of effective operators along with WCs can be eased out for a large number of models, which could be relevant for data-driven model selections based on EFT.

\section*{Acknowledgments}
 We thank Joydeep Chakrabortty for suggesting the problem and for the useful discussions. The authors also thank Kaanapuli Ramkumar for helpful suggestions. JD acknowledges the Indian Institute of Technology Kanpur (IITK) for the institutional postdoctoral fellowship grant with File No. DF/PDF/2023-IITK/2183. JD also thanks SERB, Government of India, for the national postdoctoral fellowship (NPDF) grant with File No. PDF/2023/001540.
 NA and DD acknowledge the Indian Institute of Technology Kanpur (IITK) for Institute Assistantship.

\begin{appendices}
\section{Component Green's functions }
The expressions of the first three CGFs, \(g_0(x,y), \,g_1(x,y), \,g_2(x,y)\) that we get from the Eq.~\eqref{eq:bessel} are
\begin{eqnarray} \label{app:cgf}
  g_0 (x,y) &=& \alpha \pi^{2-\frac{d}{2}} \Big[ 2^{4-d} M^{d-2} \Gamma\big(1-d/2\big) - 2^{2-d}  z^2 M^d \Gamma\big(-d/2\big) +\frac{1}{8} M^4 z^{6-d} \Gamma\big(d/2-3\big)\n \\
    & -& M^2 z^{4-d} \Gamma\big(d/2-2\big)+ 4 z^{2-d} \Gamma\big(d/2-1\big) \Big], \n \\[0.2\baselineskip]
  g_1(x,y) &=&  \alpha\pi^{2-\frac{d}{2}} \Big[ 2^{2-d}z^2 M^{d-2}\Gamma\big(1-d/2\big)-2^{4-d}M^{d-4}\Gamma\big(2-d/2\big)  +\frac{1}{4}M^2z^{6-d}\Gamma\big(d/2-3\big) \n\\
  & -& z^{4-d}\Gamma\big(d/2-2\big) \Big], \\[0.2\baselineskip]
        g_2(x,y) &=& \alpha\pi^{2-\frac{d}{2}} \Big[ -2^{1-d}z^2M^{d-4}\Gamma\big(2-d/2\big)+2^{3-d}M^{d-6}\Gamma\big(3-d/2\big) + \frac{1}{8}z^{6-d}\Gamma\big(d/2-3\big) \Big], \n
\end{eqnarray}
\\
where \(\alpha\) = \(\frac{1}{16\pi^2}\) and \(d=4-\epsilon\). 
 While computing the contribution coming from the Sunset diagram we get terms containing \(\frac{1}{z^{2a}}\) with \(a \ge 2\). At short distances \textit{i.e.,} \(z \to 0\) this terms will contribute to the \(\frac{1}{\epsilon}\) poles via the gamma function~\cite{Banerjee:2024rbc,Jack:1982hf,gel2016generalized},  
\begin{eqnarray}\label{eq:zinv_exp}
 \frac{1}{z^{2a}}=\frac{\pi^{d/2}}{4^{a-\frac{d}{2}}}\frac{\Gamma\left[ \frac{d}{2}-a \right]}{\Gamma[a]}\left(D^2\right)^n \delta^d (z)+ \mathcal{O}(\zeta^0) ,  
\end{eqnarray}
where \(a-\frac{d}{2}=n+\zeta\).
The component Green's functions at coincidence limit for order $n=2$ to $n=6$ are listed below.
\begin{eqnarray}\label{app:greenfnc}
    g_2(x,x) &=& \alpha \left[\pi^{\frac{\epsilon}{2}} 2^{\epsilon -1} M^{-\epsilon -2} \Gamma \left(\frac{\epsilon }{2}+1\right) \right],\quad
    g_3(x,x) = -\frac{\alpha }{3} \left[\pi^{\frac{\epsilon}{2}} 2^{\epsilon -1} M^{-\epsilon -4} \Gamma \left(\frac{\epsilon }{2}+2\right) \right],\nonumber\\[0.1\baselineskip]
    g_4(x,x) &=& \frac{\alpha }{3} \left[\pi^{\frac{\epsilon}{2}}  2^{\epsilon -3} M^{-\epsilon -6} \Gamma \left(\frac{\epsilon }{2}+3\right) \right],\,\,\,\,
    g_5(x,x) = -\frac{\alpha }{15} \left[\pi^{\frac{\epsilon}{2}} 2^{\epsilon -3} M^{-\epsilon -8} \Gamma \left(\frac{\epsilon }{2}+4\right) \right],\nonumber\\[0.1\baselineskip]
   g_6(x,x) &=& \frac{\alpha }{45} \left[\pi^{\frac{\epsilon}{2}} 2^{\epsilon -4} M^{-\epsilon -10} \Gamma \left(\frac{\epsilon }{2}+5\right) \right].
\end{eqnarray}
After expanding up to the power of \(\epsilon^2\), the following is a list of the expressions of the relevant gamma functions containing the pole. 
\begin{eqnarray}
 &&\Gamma\left(\frac{\epsilon}{2}-1\right) = -\frac{2}{\epsilon}+\gamma-1+  \frac{\epsilon}{24}(-12+ 12\gamma -6\gamma^2 -\pi^2)\n\\[-0.1\baselineskip] 
 &&\hspace{19mm}+\frac{\epsilon^2}{48}\left(-12+12\gamma-6\gamma^2 +2\gamma^3 -\pi^2 + \gamma\pi^2-2\psi^{(1)}(2)\right),\n \\[0.25\baselineskip]
&&\Gamma\left(\frac{\epsilon}{2}\right) = \frac{2}{\epsilon} - \gamma +\frac{\epsilon}{24}(6\gamma^2+ \pi^2) +\frac{\epsilon^2}{24}\left(-\gamma^3-\frac{\gamma\pi^2}{2}+\psi^{(1)}(2) \right),\nonumber\\
&&\Gamma\left(1+\frac{\epsilon}{2}\right) \,\,= 1-\frac{\gamma\epsilon}{2}+\frac{\epsilon^2}{48}\left(\pi^2 +6\gamma^2\right),
\end{eqnarray}
where \(\psi^{(n)}(x)\) is \(n\)-th derivative of digamma function \(\psi(x)\), and $\gamma$ is the universal Euler–Mascheroni constant.  

\subsection{An extra contribution coming from the finite part of \texorpdfstring{$g_0^2(x-y)$}{Lg} }\label{app:finite}
    In the expansion of $\dfrac{1}{z^{2a}}$ in the Eq.~\eqref{eq:zinv_exp}, there's a finite part, that we didn't consider in our calculation. That can be calculated by doing Fourier transform~\cite{Jack:1982hf,gel2016generalized} as 
    \begin{eqnarray}
        \int \dfrac{1}{|z|^{2a}} e^{ikz} d^d z = \pi^{d/2} \dfrac{\Gamma(d/2 - a)}{\Gamma(a)} \left(\dfrac{1}{4} k^2 \right)^{a - \frac{d}{2}}.
    \end{eqnarray}
    While calculating $g_0^2(x-y)$, in which case $a = 2 - \epsilon$ and $d = 4 - \epsilon$, we get 
    \begin{eqnarray}
        \pi^{2 -\epsilon/2} \dfrac{\Gamma(\epsilon/2)}{\Gamma( 2 - \epsilon)} \bigg(\dfrac{1}{4} k^2 \bigg)^{- \epsilon/2} \!\!
        &=& \pi^{2 -\epsilon/2} \dfrac{\Gamma(\epsilon/2)}{\Gamma( 2 - \epsilon)} \exp\left(\! \! -\dfrac{\epsilon}{2} \log\left(\dfrac{k^2}{4}\right) \right)\n \\
        &=& \pi^{2 -\epsilon/2} \dfrac{\Gamma(\epsilon/2)}{\Gamma( 2 - \epsilon)} \bigg(1 \! \! -\dfrac{\epsilon}{2} \log\left(\dfrac{k^2}{4}\right) +\mathcal{O}\left(\epsilon^2\right) \bigg).
    \end{eqnarray}
As $\Gamma(\epsilon/2) \sim \dfrac{2}{\epsilon} $, it is clear that we get a finite piece, which is $ \! \left(-\pi^2 \log\dfrac{k^2}{4} \right)$. In the configuration space, this looks like 
\begin{eqnarray}
    \int \log \left(k^2\right) e^{ikz} d^d k = \log\left(-D^2\right) \delta^d(z).
\end{eqnarray}
After applying $\overline{MS}$ regularization scheme, we get 
$ \left.g_0^2(x-y) \right\vert_{\rm finite} = - \alpha \log\left(\! \!-\dfrac{D^2}{\mu^2}\right) \delta^4(x-y)$.
So, in the two-loop effective Lagrangian, the additional contribution due to this part is 
\begin{eqnarray}
    \mathcal{L}_{(2)} &\supset& -\dfrac{1}{12} \text{Tr}\Big[ \int d^d x d^d y V_{(3)}(x)\bigg(  \! \! -3 \alpha \log\left(\!  -\frac{D^2}{\mu^2}\right) g_1(x, y)\Tilde{b}_0(x,y)^2 \Tilde{b}_1(x, y)  \bigg) \n \\
    & \times& V_{(3)}(y) \delta^4(x-y) \Big].
\end{eqnarray}

 \section{The poles and finite parts of the three distinct vacuum diagrams}
\paragraph{Sunset diagram:}For the sunset diagram, the coefficients of $1/\epsilon^2,\, 1/\epsilon$, and the finite parts that appear in the Lagrangian are
\begin{eqnarray}\label{app:a1}
\mathcal{C}_{(2)}^a|_{\epsilon^{-2}}&=&\text{tr} \bigg[\frac{1}{2} V_{(3)}^2 \Big(\alpha ^2 \tilde{b}_1\tilde{b}_0^2+\alpha ^2 \tilde{b}_0^3 M^2\Big) \bigg],\\
\n\\
 \mathcal{C}_{(2)}^a|_{\epsilon^{-1} }&=& \text{tr} \bigg[ \frac{1}{24} \alpha ^2 \tilde{b}_0^3 V_{(3)} D^2 V_{(3)} + \frac{\alpha^2}{12} V_{(3)}^2 \bigg\{ \tilde{b}_0^3 \bigg(9 M^2 
     -6 M^2 \log (\frac{M^2}{4 \pi e^{-\gamma}}) \bigg)\nonumber\\
    & -&3  \tilde{b}_1\tilde{b}_0^2 \left( 2 \log (\frac{M^2}{4 \pi e^{-\gamma}}) 
     -1\right)-6   F \tilde{b}_0^2\bigg\} \,  \bigg],\\
     \n\\
        \mathcal{C}_{(2)}^a|_{\epsilon^0}\,\,\, & =& \text{tr} \Bigg[ \frac{\alpha ^2 \tilde{b}_0^3}{24} V_{(3)} D^2 V_{(3)} \left( \frac{13}{4} + \log (4 \pi e^{- \gamma} )\right) + \frac{1}{96} \alpha ^2 \tilde{b}_0^2 V_3^2 \bigg[\tilde{b}_0 \bigg\{2M^2 \n\\[0.08\baselineskip]
        &\times& \Big(12 (\gamma -3) \gamma +\pi ^2+30\Big)+24 M^2 \Big(\log (4 \pi ) (-2 \gamma -4 \log (M)+3)\n\\
        & +&2 \log (M) (2 \gamma +\log (M)-3)+\log ^2(4 \pi )\Big)\bigg\}+2 \tilde{b}_1\Big(12 (\gamma -1) \gamma \n\\
        & +&12 \Big[-2 \log (4 \pi ) (\gamma +2 \log (M))+2 \log (M) (2 \gamma +\log (M)-2)\n\\
        & +&\log ^2(4 \pi )\Big]+\pi ^2-6+12 \log (4 \pi )\Big)+24 F (\gamma -2-\log (4 \pi ))\bigg] \, \Bigg].\label{app:a2}
    \end{eqnarray}
\paragraph{Infinity diagram:}In the context of the infinity diagram, the coefficients of $1/\epsilon^2,\, 1/\epsilon$, and the finite parts  appearing in the Lagrangian are
\begin{eqnarray}\label{app:b1}
\mathcal{C}_{(2)}^b|_{\epsilon^{-2}} & =& \dfrac{1}{2} \text{tr} \bigg[ \alpha^2 V_{(4)} \Big(  \tilde{b}_1+  \tilde{b}_0 M^2\Big)^2 \bigg],\\
  \n\\
 \mathcal{C}_{(2)}^b|_{\epsilon^{-1}} & =& -\frac{1}{2} \text{tr} \bigg[ \alpha^2  V_{(4)} \left(\tilde{b}_0 M^2+\tilde{b}_1\right) \bigg\{ \tilde{b}_0 M^2 \left( \log (\frac{M^2}{4\pi e^{-\gamma}})-1\right) \nonumber\\ 
        & + & \tilde{b}_1\log (\frac{M^2}{4\pi e^{-\gamma}})+F\bigg\} \, \bigg] , \\
\n\\
        \mathcal{C}_{(2)}^b|_{\epsilon^0} \,\,\,& =& \frac{1}{48} \text{tr} \Bigg[  \alpha ^2 V_4 \bigg[6 \bigg\{\tilde{b}_0 M^2 \{\log (\frac{M^2}{4\pi e^{-\gamma}})-1\}+\tilde{b}_1\log (\frac{M^2}{4\pi e^{-\gamma}})+F\bigg\}^2\n\\[0.1\baselineskip]
        & +&\left(\tilde{b}_0 M^2+\tilde{b}_1\right) \bigg\{\tilde{b}_0 M^2 \bigg(6 \Big((\gamma -2) \gamma +2+4 \log ^2(2)+\log ^2(\pi )\n\\[0.1\baselineskip]
        & +&\log (16)+(2+\log (16)) \log (\pi )\Big)+12 \log (M^2) (\gamma +\frac{1}{2}\log (M^2)-1)\n\\[0.1\baselineskip]
        & -&12 \log (4 \pi ) (\gamma + \log (M^2))+\pi ^2\bigg)+\tilde{b}_1\Big(6 \gamma ^2+6 [\log (4 \pi )- \log (M^2)]\n\\
        & \times& [-2 \gamma -\log (M^2)+\log (4\pi )]+\pi ^2\Big)\bigg\}\bigg] \, \Bigg].\label{app:b2}
    \end{eqnarray}
\paragraph{Counter-term diagram:} For the counter-term diagram, the coefficients of $1/\epsilon^2,\, 1/\epsilon$, and the finite parts appearing in the Lagrangian are
\begin{eqnarray}\label{app:ct1}
        \mathcal{C}_{(2)}^{\rm ct}|_{\epsilon^{-2}} & =& -\frac{1}{2} \alpha^2 \text{tr} \Big[ \left(\tilde{b}_0 M^2+\tilde{b}_1\right) \left(M^4 \tilde{b}_0^{''}+2 M^2 \tilde{b}_1^{''}+ \tilde{b}_2^{''}\right) \Big],\\
        \n\\
     \mathcal{C}_{(2)}^{\rm ct}|_{\epsilon^{-1}} & =& \frac{1}{4} \alpha^2 \text{tr} \bigg[ \left(M^4 \tilde{b}_0^{''}+2 M^2 \tilde{b}_1^{''}+ \tilde{b}_2^{''}\right) \bigg\{  \tilde{b}_0 M^2 \left(\log (\frac{M^2}{4 \pi e^{\gamma}})-1\right)  \n\\ 
        & + & \tilde{b}_1\log (\frac{M^2}{4 \pi e^{\gamma}})+F\bigg\} \, \bigg],\\
        \n\\
        \mathcal{C}_{(2)}^{\rm ct}|_{\epsilon^0}\label{app:ct2}\,\,\,
        & =& -\frac{1}{96} \alpha^2 \text{tr} \Bigg[ \left(M^4 \tilde{b}_0^{''}+2 M^2 \tilde{b}_1^{''}+\tilde{b}_2^{''}\right) \bigg(\tilde{b}_0 M^2 \Big(6 \{(\gamma -2) \gamma +2 \n\\[0.1\baselineskip]
        & +& 4 \log ^2(2)  +\log ^2(\pi )+\log (16)+(2+\log (16)) \log (\pi )\}+12 \log (M^2) \n\\[0.15\baselineskip]
        & \cross& (\gamma +\frac{1}{2}\log (M^2)-1)-12 \log (4 \pi ) (\gamma + \log (M^2))+\pi ^2\Big)\\
        & +&\tilde{b}_1\Big(6 \gamma ^2  +6 (\log (4 \pi )- \log (M^2)) (-2 \gamma -\log (M^2)+\log (4\pi )) 
         +\pi ^2\Big)\bigg) \Bigg].\n
\end{eqnarray}
\paragraph{Resultant contributions:}
The coefficients of $1/\epsilon^2,\, 1/\epsilon$, and the finite parts of the total contribution, obtained by summing the three diagrams, can be expressed as
\begin{eqnarray}\label{app:tot1}
    \mathcal{C}_{(2)}|_{\epsilon^{-2}} & =&-\frac{1}{2} \alpha ^2 \text{tr} \Big[ \left(\tilde{b}_0 M^2+\tilde{b}_1\right) \Big(M^4 \tilde{b}_0^{''}-\tilde{b}_0 M^2 V_{(4)}+2 M^2 \tilde{b}_1^{''}-\tilde{b}_0^2 V_{(3)}^2\n\\
    & -&\tilde{b}_1V_{(4)}+\tilde{b}_2^{''}\Big) \Big],\\
    \n\\
    \mathcal{C}_{(2)}|_{\epsilon^{-1}} \label{app:tot2} & =& \frac{1}{24} \alpha ^2 \text{tr} \Bigg[ \tilde{b}_0^3\, V_{(3)} D^2 V_{(3)} + \tilde{b}_0^2\, V_3^2 \bigg\{6\tilde{b}_0 M^2 \bigg(3-2 \log (\frac{M^2}{4 \pi e^{-\gamma}} )\bigg)\n\\[0.1\baselineskip]
    & + &6 \tilde{b}_1\left(1-2 \log (\frac{M^2}{4 \pi e^{-\gamma}} )\right)-12 F\bigg\}-12 V_4 \left(\tilde{b}_0 M^2 + \tilde{b}_1\right) \bigg(\tilde{b}_0 M^2 \n\\[0.1\baselineskip]
    & \times & \Big[\log (\frac{M^2}{4 \pi e^{-\gamma}} )-1\Big] + \tilde{b}_1\log (\frac{M^2}{4 \pi e^{-\gamma}} )+F\bigg)+6 (M^4 \tilde{b}_0^{''} + 2 M^2 \tilde{b}_1^{''} \! \n\\[0.1\baselineskip]
    & + & \tilde{b}_2^{''}) \Big\{\tilde{b}_0 M^2 \Big[\log (\frac{M^2}{4 \pi e^{-\gamma}} )-1\Big]+\tilde{b}_1\log (\frac{M^2}{4 \pi e^{-\gamma}} )+F\Big\} \Bigg],\\
    \n \\
    \text{\large $ \mathcal{C}_{(2)}|_{\epsilon^0}$}\label{app:tot3}\,\,\,
    & =& \frac{1}{24} \alpha ^2 \text{tr} \Bigg[  \tilde{b}_0^3 V_{(3)} D^2 V_{(3)} \Big( \frac{13}{4} + \log (4 \pi e^{- \gamma} )\Big) + \frac{1}{4} \tilde{b}_0^2 V_3^2 \bigg\{\tilde{b}_0 \bigg(2 M^2 ( 30 \n\\[0.1\baselineskip]
    & +& 12\gamma (\gamma -3) +\pi ^2)+24 M^2 \Big\{\log (4 \pi ) (-2 \gamma +3 -2 \log (M^2))\n\\[0.1\baselineskip]
    & +& \log (M^2) (2 \gamma -3 +\frac{1}{2}\log (M^2))+\log ^2(4 \pi )\Big\}\bigg)+2 \tilde{b}_1\bigg(12 \gamma (\gamma -1) \n\\[0.1\baselineskip]
    & +&12 \Big\{-2 \log (4 \pi ) (\gamma + \log (M^2))+ \log (M^2) (2 \gamma -2 +\frac{1}{2}\log (M^2))\n\\[0.1\baselineskip]
    & +&\log ^2(4 \pi )\Big\}+\pi ^2-6+12 \log (4 \pi )\bigg)+24 F (\gamma -2-\log (4 \pi ))\bigg\}\n\\[0.1\baselineskip]
    & +&2 V_4 \bigg\{6 \Big(\tilde{b}_0 M^2 \Big[\log (\frac{M^2}{4 \pi e^{-\gamma}} )-1\Big]+\tilde{b}_1\log (\frac{M^2}{4 \pi e^{-\gamma}} )+F\Big)^2\n\\[0.1\baselineskip]
    & +&\left(\tilde{b}_0 M^2 + \tilde b_1\right) \Big(\tilde{b}_0 M^2 \Big[6 (\gamma -2) \gamma +12 \log (M^2) (\gamma -1 +\frac{1}{2}\log (M^2))\n\\[0.15\baselineskip]
    & -&12 \log (4 \pi ) \{\gamma +2 \log (M)\}+\pi ^2+6 \{2+4 \log ^2(2)+\log ^2(\pi )\n\\[0.15\baselineskip]
    & +&\log (16)+(2+\log (16)) \log (\pi )\}\Big]+\tilde{b}_1\Big[6 \gamma ^2+6 \{\log (4 \pi )- \log (M^2)\} \n\\[0.1\baselineskip]
    & \times& \{-2 \gamma -\log (M^2)+\log (4\pi )\}+\pi ^2\Big]\Big)\bigg\}-\left(M^4 \tilde{b}_0^{''}+2 M^2 \tilde{b}_1^{''}+\tilde{b}_2^{''}\right) \n\\[0.1\baselineskip]
    & \times& \bigg\{\tilde{b}_0 M^2 \bigg(6 (\gamma -2) \gamma +12 \log (M^2) (\gamma -1 +\frac{1}{2}\log (M^2))-12 \log (4 \pi )\n\\[0.1\baselineskip]
    & \times& (\gamma +2 \log (M))+\pi ^2+6 \big[2+4 \log ^2(2)+\log ^2(\pi )+\log (16)\n \\[0.1\baselineskip]
    &+& (2 +\log (16)) \log (\pi )]\bigg)+\tilde{b}_1\Big(6 \gamma ^2+6 [\log (4 \pi )- \log (M^2)\big] \big[-2 \gamma\n \\
    & -&\log (M^2)+\log (4\pi )\big]+\pi ^2\Big)\bigg\} \Bigg].
\end{eqnarray}
\paragraph{Considering the explicit value of $F$:}By substituting the explicit value of \( F \), the coefficients of $1/\epsilon^2,\, 1/\epsilon$, and the finite parts of the total contributions can be expressed as
\begin{eqnarray}
         \mathcal{C}^\prime_{(2)}|_{\epsilon^0}\label{app:F1}\,\,\,
         & =& \frac{1}{24} \alpha^2 \text{tr} \Bigg[  \tilde{b}_0^3 V_{(3)} D^2 V_{(3)} \Big( \frac{13}{4} + \log (4 \pi e^{- \gamma} )\Big) + \frac{1}{1200}\bigg[ 60 \tilde{b}_0^2 V_3^2 \bigg\{5 \tilde{b}_0 \bigg(2 [12 (\gamma -3) \gamma \n\\[0.1\baselineskip]
        & +&\pi ^2+30] M^2+24 M^2 \Big\{\log (4 \pi ) (-2 \gamma -2 \log (M^2)+3)+ \log (M^2) (2 \gamma -3 \n\\[0.1\baselineskip]
        & +&\frac{1}{2}\log (M^2))+\log ^2(4 \pi )\Big\}\bigg)+\frac{2}{M^{10}} \Big\{-3 \tilde{b}_5 M^2+5 M^4 (6 \tilde{b}_2 M^4-2 \tilde{b}_3 M^2 + \tilde{b}_4) \n\\[0.1\baselineskip]
        & +& 2 \tilde{b}_6\Big\}(\gamma -2-\log (4 \pi ))+10 \tilde{b}_1\Big\{12 (\gamma -1) \gamma +12 \Big(-2 \log (4 \pi ) (\gamma + \log (M^2)) \n \\[0.1\baselineskip]
        & +& \log (M^2) (2 \gamma +\frac{1}{2}\log (M^2)-2)+\log ^2(4 \pi )\Big)+\pi ^2-6+12 \log (4 \pi )\Big\}\bigg\}\n\\[0.1\baselineskip]
        & -&300 \left(M^4 \tilde{b}_0^{''}+2 M^2 \tilde{b}_1^{''}+\tilde{b}_2^{''}\right) \bigg\{\tilde{b}_0 M^2 \Big(6 \gamma (\gamma -2) +12 \log (M^2) (\gamma -1 +\frac{1}{2}\log (M^2))\n\\[0.1\baselineskip]
        & -& 12 \log (4 \pi ) (\gamma + \log (M^2))+\pi ^2+6 \big[2+4 \log ^2(2)+\log (16) (1+\log (\pi )) + \log (\pi ) \n\\[0.2\baselineskip]
        & \times& (2+\log (\pi ))\big]\Big)+\tilde{b}_1\Big[6 \gamma ^2+6 (\log (4 \pi )- \log (M^2)) (-2 \gamma -\log (M^2)+\log (4\pi ))\n\\[0.1\baselineskip]
        & +&\pi ^2\Big]\bigg\}+V_4 \bigg\{600 \Big(\tilde{b}_0 M^2+\tilde{b}_1\Big) \Big(\tilde{b}_0 M^2 \Big[6 (\gamma -2) \gamma +12 \log (M^2) (\gamma -1 +\frac{1}{2}\log (M^2))\n\\[0.2\baselineskip]
        & -&12 \log (4 \pi ) (\gamma +2 \log (M))+\pi ^2+6 \big(2+4 \log ^2(2)+\log (16) (1+\log (\pi )) \n\\[0.2\baselineskip]
        & +&\log (\pi ) (2+\log (\pi ))\big)\Big]+\tilde{b}_1\big[6 \gamma ^2+6 (\log (4 \pi )- \log (M^2)) (-2 \gamma -\log (M^2)+\log (4\pi ))\n\\[0.1\baselineskip]
        & +&\pi ^2\big]\Big)+\frac{1}{M^{20}}\bigg(\!-3 \tilde{b}_5 M^2+5 M^4 \Big[-2 \tilde{b}_3 M^2+6 M^4 \Big(2 \tilde{b}_0 M^4 (\gamma + \log (M^2)-1\n\\[0.1\baselineskip]
        & -&\log (4 \pi ))+2 \tilde{b}_1M^2 (\gamma + \log (M^2)-\log (4 \pi ))+\tilde{b}_2\Big)+\tilde{b}_4\Big]+2 \tilde{b}_6\bigg)^2\bigg\} \bigg] \Bigg],\\
   \n\\
        \mathcal{C}^\prime_{(2)}|_{\epsilon^{-1}}\label{app:F2}
        & =& \frac{1}{24} \alpha ^2 \text{tr} \Bigg[ \tilde{b}_0^3 V_{(3)} D^2 V_{(3)} + 2 \tilde{b}_0^2 V_3^2 \bigg\{5 \tilde{b}_0 M^{10} \Big(6 M^2 (-2 \gamma +3+2 \log (4 \pi ))\n\\
        & -&12 M^2 \log (M^2)\Big)+3 \tilde{b}_5 M^2-5 M^4 \Big(6 \tilde{b}_1M^6 (2 \gamma -1 +2 \log (M^2)-2 \log (4 \pi ))\n\\
        & +&6 \tilde{b}_2 M^4-2 \tilde{b}_3 M^2+\tilde{b}_4\Big)-2 \tilde{b}_6\bigg\}-\left(M^4 \tilde{b}_0^{''}+2 M^2 \tilde{b}_1^{''}+\tilde{b}_2^{''}\right) \Big[3 \tilde{b}_5 M^2\n\\
        & -&5 M^4 \Big(-2 \tilde{b}_3 M^2+6 M^4 \Big(2 \tilde{b}_0 M^4 (\gamma -1 +\log (M^2)-\log (4 \pi ))+2 \tilde{b}_1M^2 (\gamma \n\\
        & +& \log (M^2)-\log (4 \pi ))+\tilde{b}_2\Big)+\tilde{b}_4\Big)-2 \tilde{b}_6\Big]+2 V_4 \left(\tilde{b}_0 M^2+\tilde{b}_1\right) \bigg\{3 \tilde{b}_5 M^2\n\\
        & -&5 M^4 \Big(-2 \tilde{b}_3 M^2+6 M^4 [2 \tilde{b}_0 M^4 (\gamma -1 + \log (M^2)-\log (4 \pi ))+2 \tilde{b}_1M^2 (\gamma \n\\
        & +& \log (M^2)-\log (4 \pi ))+\tilde{b}_2]+\tilde{b}_4\Big)-2 \tilde{b}_6\bigg\} \Bigg].
\end{eqnarray}
\section{The rest of the coefficients of the Lagrangian }\label{app:coeff}
The expressions of the remaining coefficients of the Lagrangian given in Eq.~\eqref{eq:effLag}, such as $\mathcal{C}_{-4}, \mathcal{C}_{-6}$, and so on, are provided below. Note that these coefficients may include terms with dimensions more than six, but we only consider terms with dimensions up to six, assuming $U$ has a minimum operator dimension of one.
\begin{eqnarray}
    \mathcal{C}_{-4} &=&  \frac{1}{24} U_{\phi }^2 U^3 \Big[2+\log (4 \pi e^{-\gamma})\Big]+\frac{1}{48} G_{\mu \nu }^2 U_{\phi }^2 U \Big[2+\log (4 \pi e^{-\gamma})\Big] + \frac{1}{360} G_{\mu \nu } G_{\nu \rho } G_{\rho \mu } U_{\phi }^2 \n \\
    & \times& \Big[2+\log (4 \pi e^{-\gamma})\Big]- \frac{1}{240} J_{\nu }^2 U_{\phi }^2 \Big[2+\log (4 \pi e^{-\gamma})\Big]- \frac{1}{48} U_{\mu }^2 U_{\phi }^2 \Big[2+\log (4 \pi e^{-\gamma})\Big] \n \\
    & +& \frac{1}{96} U^4 U_{\phi \phi } \left[1-2 \log (\frac{M^2}{4\pi e^{-\gamma}})\right] - \frac{1}{240}  (U G_{\mu \nu })^2 U_{\phi \phi } \left[1-\log (\frac{M^2}{4\pi e^{-\gamma}})\right] \n \\
    & -& \frac{1}{480}  U^2 G_{\mu \nu }^2 U_{\phi \phi } \left[3+2 \log (\frac{M^2}{4\pi e^{-\gamma}})\right]
    + \frac{1}{48}  U U_{\mu }^2 U_{\phi \phi } \log (\frac{M^2}{4\pi e^{-\gamma}}) \n \\
    & + & \frac{1}{48}  U^2 U_{\mu \mu } U_{\phi \phi } \log (\frac{M^2}{4\pi e^{-\gamma}})
     - \frac{1}{1440} U_{\mu \mu }^2 U_{\phi \phi } \left[1-6 \log (\frac{M^2}{4\pi e^{-\gamma}})\right] \n \\
     & + & \frac{1}{120}  U J_{\nu } U_{\nu } U_{\phi \phi } \left[1-\log (\frac{M^2}{4\pi e^{-\gamma}})\right],\\[\baselineskip]
    \mathcal{C}_{-6} &=& -\frac{1}{48} U_{\phi }^2 U^4 \Big[2+\log (4 \pi e^{-\gamma})\Big] - \frac{1}{240} U_{\phi }^2 (U G_{\mu \nu })^2 \Big[2+\log (4 \pi e^{-\gamma})\Big]  - \frac{1}{60} G_{\mu \nu }^2 U_{\phi }^2 U^2 \n \\
    & \times& \Big[2+\log (4 \pi e^{-\gamma})\Big]  - \frac{1}{48} U_{\mu \mu } U_{\phi }^2 U^2 \Big[2+\log (4 \pi e^{-\gamma})\Big] - \frac{1}{240} U_{\mu \mu }^2 U_{\phi }^2 \Big[2+\log (4 \pi e^{-\gamma})\Big] \n \\
    &  +& \frac{1}{120} J_{\nu } U_{\nu } U_{\phi }^2 U \Big[2+\log (4 \pi e^{-\gamma})\Big]  - \frac{1}{120} U^5 U_{\phi \phi } \left[1-\log (\frac{M^2}{4\pi e^{-\gamma}})\right] \n \\
    & + &\frac{1}{480} U^2 U_{\mu }^2 U_{\phi \phi } \left[11-6\log (\frac{M^2}{4\pi e^{-\gamma}})\right] + \frac{1}{720}  U^3 U_{\mu \mu } U_{\phi \phi } \left[13-3\log (\frac{M^2}{4\pi e^{-\gamma}})\right] , \\[\baselineskip]
        \mathcal{C}_{-8} &=& \frac{1}{80} U_{\phi }^2 U^5 \Big[2+\log (4 \pi e^{-\gamma})\Big]   + \frac{1}{80} U_{\mu }^2 U_{\phi }^2 U^2 \Big[2+\log (4 \pi e^{-\gamma})\Big]  + \frac{1}{40} U_{\mu \mu } U_{\phi }^2 U^3 \n \\
      & \times& \Big[2+\log (4 \pi e^{-\gamma})\Big]  + \frac{1}{720} U^6 U_{\phi \phi } \left[4-3 \log (\frac{M^2}{4\pi e^{-\gamma}})\right], \\[\baselineskip]
      \mathcal{C}_{-10} &=& -\frac{1}{120}  U^6 U_{\phi }^2\Big[2+\log (4 \pi e^{-\gamma})\Big].
\end{eqnarray}
Even if $U$ has the lowest operator dimension of one, the other coefficients of the effective Lagrangian given in Eq.~\eqref{eq:effLag}, such as $\mathcal{C}_{-12}, \,\mathcal{C}_{-14}$, and so on, do not contribute operators up to dimension six. Therefore, we do not consider them here.
\section{One-loop dimension six operators for triple and doublet models} \label{app:models}
\subsection{Electroweak triplet with hypercharge \texorpdfstring{$Y_\Delta=1$}{Lg}}\label{app:onelooptrp}
For the electroweak triplet model, the following are dimension six bosonic operator structures at the one-loop level.
\begin{eqnarray}
        \dfrac{1}{2} \text{tr} \big[U \big]\label{app:dim1}  
        &=& \dfrac{1}{2} \Big[\left(U_{11}\right)_{ii}+\left(U_{22}\right)_{ii}\Big]\n\\ 
        & =& \dfrac{3}{2} \lambda_1 \lvert H \rvert ^2 + \left(2 \lambda_2 + \dfrac{3}{2} \lambda_3\right) \lvert \Delta_c \rvert ^2 + \dfrac{3}{4} \lambda_4 \lvert H \rvert ^2 \n \\
        & \supset& \dfrac{\mu^2_\Delta}{m^6}\left(2 \lambda_2 + \dfrac{3}{2} \lambda_3\right)\Big[2 \mathcal{O}_H - \left(\dfrac{\lambda_1}{2}+\dfrac{\lambda_4}{4}\right) \mathcal{O}_6 \Big],\\
  \n\\
        \dfrac{1}{2} \text{tr} \big[U^3\big]  
        &=& \dfrac{1}{2} \Big[\left(U_{11}^3\right)_{ii}+\left(U_{22}^3\right)_{ii}\Big]\n\\
        &\supset& \dfrac{3}{64} \Big[\, 8 \lambda^3_1 + 3\lambda^3_4 + 12 \lambda^2_1 \lambda_4 + 10 \lambda_1 \lambda^2_4 \, \Big] \mathcal{O}_6,\\
   \n\\[0.4\baselineskip]
        \dfrac{1}{2}\text{tr}\big[U_\mu^2\big] 
        &=& \dfrac{1}{2} \Big[\left(D_\mu U_{11}\right)_{ij}\left(D^\mu U_{11}\right)_{ji}+\left(D_\mu U_{22}\right)_{ij}\left(D^\mu U_{22}\right)_{ji}\Big]\n\\
        &\supset& \dfrac{3}{8}\Big [(4 \lambda^2_1 + \lambda^2_4 + 4\lambda_1 \lambda_4)\mathcal{O}_H + \dfrac{2}{3} \lambda^2_4(\mathcal{O}_T + 2 \mathcal{O}_R) \Big],\\
    \n \\[0.3\baselineskip]
        \dfrac{1}{2} \text{tr} \big[U G_{\mu \nu}^2\big] 
        &=& \dfrac{1}{2} \Big[\left(U_{11}\right)_{ij}\left(G_{\mu \nu}^2\right)_{ji}+\left(U_{22}\right)_{ij}\left(G_{\mu \nu}^2\right)_{ji}\Big]\n\\
        &\supset& - \left(\dfrac{\lambda_1}{2} + \dfrac{\lambda_4}{4}\right)\big(2 \mathcal{O}_{WW} + 3  \mathcal{O}_{BB} \big) + \lambda_4 \mathcal{O}_{WB},\\
   \n \\[0.3\baselineskip]
        \dfrac{1}{2} \text{tr} \big[J_\nu^2\big] 
        &=& - 4 g^2_W \mathcal{O}_{2W} - 6 g^2_Y \mathcal{O}_{2B},\quad
       \dfrac{1}{2} \text{tr} \big[G_{\mu \nu}^3\big] = - 6 g^2_W \mathcal{O}_{3W}. 
\end{eqnarray}
\subsection{Electroweak doublet with hypercharge \texorpdfstring{$Y_\Phi=-1/2$}{Lg}}\label{app:oneloop2HDM}
For the two Higgs doublet model, dimension six bosonic operator structures at the one-loop level are provided below.
\begin{eqnarray}
        \dfrac{1}{2} \text{tr}  \big[U\big] 
        &=& \dfrac{3}{2} \lambda_\Phi \Phi^\dagger_c \Phi_c - 3 \eta_\Phi ( \tilde{H}^\dagger \Phi_c + \Phi^\dagger_c \tilde{H}) + \left(2 \lambda _1 +  \lambda _2\right)|H|^2 \n\\
        &\supset& \dfrac{3}{2 m^4_\Phi } \lambda_\Phi \eta^2_H \mathcal{O}_6 - 6 \eta_\Phi \eta_H \bigg[ \dfrac{1}{m^4_\Phi} (\mathcal{O}_H + \mathcal{O}_R) - \dfrac{\lambda_1 + \lambda_2}{m^4_\Phi} \mathcal{O}_6 \bigg], \\
  \n\\[0.1\baselineskip]
        \dfrac{1}{2} \text{tr} \big[U^3\big] 
        &=& \dfrac{1}{2}  \Big[ (U^3_{11})_{ii} + (U^3_{22})_{ii} + 3(U_{11} U_{12} U_{21} + U_{22} U_{12} U_{21})_{ii} \Big]  \\
        &\supset& \Big[  2\lambda_1^3 + 3\lambda_1^2 \lambda_2 + 3 \lambda_1\lambda_2^2 + \lambda_2^3 + 12 \lambda_3^2 (\lambda_1 + \lambda_2) \Big] \mathcal{O}_6, \n \\
   \n\\[0.1\baselineskip]
        \dfrac{1}{2}\text{tr}\big[U_\mu^2\big] 
        &=& \dfrac{1}{2} \Big[\left(D_\mu U_{11}\right)_{ij}\left(D^\mu U_{11}\right)_{ji} + \left(D_\mu U_{22}\right)_{ij}\left(D^\mu U_{22}\right)_{ji}  + 2 \left( D_\mu U_{12} \right)_{ij} \left( D_\mu U_{21} \right)_{ji} \Big]\n \\
        &\supset& \Big[ (4\lambda_1^2 + \lambda_2^2 + 4\lambda_3^2 + 4\lambda_1 \lambda_2) \mathcal{O}_H + (\lambda_2^2 - 4 \lambda_3^2) \mathcal{O}_T + 2(\lambda
        _2^2 + 4 \lambda_3^2)\mathcal{O}_R \Big],\\
    \n \\[0.1\baselineskip]
        \dfrac{1}{2} \text{tr} \big[U G_{\mu \nu}^2\big] 
        &=& \dfrac{1}{2} \Big[\left(U_{11}\right)_{ij}\left(G_{\mu \nu}^2\right)_{ji}+\left(U_{22}\right)_{ij}\left(G_{\mu \nu}^2\right)_{ji}\Big]\n\\
        &\supset& -(2\lambda_1 + \lambda_2)\left(\dfrac{1}{4} \mathcal{O}_{WW} + \dfrac{1}{4} \mathcal{O}_{BB}\right) - \dfrac{1}{2} \lambda_2 \mathcal{O}_{WB},\\
    \n \\[0.1\baselineskip]
        \frac{1}{2}\text{tr}\big[J_\nu^2\big]\label{app:dim2} 
        &=&\frac{1}{2} \text{tr}(D_\mu G_{\mu \nu})^2 = - g^2_W \mathcal{O}_{2W} - g^2_Y \mathcal{O}_{2B}, \quad
        \frac{1}{2}\text{tr}\big[ G_{\mu\nu}^3\big] = - \dfrac{3}{2} g_W^2 \mathcal{O}_{3W}.
\end{eqnarray}
\subsection{Some useful relations and definitions}
To write the potential in the form of a $6\times 6$ matrix from the trace part of the Lagrangian (see Eq.~\eqref{eq:trptpot}), we present here some algebraic relations for the electroweak triplet model.
\begin{eqnarray}\label{app:trpttrace1}
    \text{Tr}[\Delta^\dagger \Delta]\quad \,
    &=& \text{Tr}[\Delta^*_i \Delta_j \tau^i \tau^j] = \dfrac{1}{2}\Delta^*_i \Delta_i = \, \dfrac{1}{2}|\Delta|^2 ,\\
        \text{Tr}\left[\left(\Delta^\dagger \Delta\right)^2\right] \label{app:trpttrace2}
        & =& \Delta^*_i \Delta_j \Delta^*_k \Delta_l \text{Tr} [\tau^i \tau^j \tau^k \tau^l]  
         = \dfrac{1}{8} \Delta^*_i \Delta_j \Delta^*_k \Delta_l (\delta_{ij} \delta_{kl} + \delta_{il} \delta_{jk} - \delta_{ik} \delta_{jl})\n\\
        & =& \dfrac{1}{4} (\Delta^*_i \Delta_i)(\Delta^*_j \Delta_j) -\dfrac{1}{8}(\Delta^*_i \Delta^*_i)(\Delta_j \Delta_j)
        = \dfrac{1}{4}|\Delta|^4  -\dfrac{1}{8}(\Delta^\ast)^2(\Delta)^2.
\end{eqnarray}
We present some algebraic relations for both scenarios to construct the dimension six effective operators, which are given below. 
\begin{align}\label{app:essential1}
       & \big( H^\dagger\tau^a H\big)P^2 \big( H^\dagger\tau^a H\big)
        = D_\mu\big( H^\dagger\tau^a H\big)D^\mu \big( H^\dagger\tau^a H\big)
        = \frac{1}{2}\big(\mathcal{O}_T+2\mathcal{O}_R\big),\\[0.45\baselineskip]
       & \big( \tilde{H}^\dagger\tau^a H\big)P^2 \big( H^\dagger\tau^a \tilde{H}\big)
        =D_\mu\big( \tilde{H}^\dagger\tau^a H\big)D^\mu \big( H^\dagger\tau^a \tilde{H}\big) 
        = \mathcal{O}_H , \\[0.45\baselineskip]
 &\big(H^\dag D_\mu H \big)^2 + \big( (D_\mu H)^\dag H \big) = \mathcal{O}_T + \mathcal{O}_H, \text{\hspace{5cm}} \\[0.45\baselineskip]
 &\big( H^\dagger H\big)P^2 \big( H^\dagger H\big)
        = D_\mu\big( H^\dagger H\big)D^\mu \big( H^\dagger H\big)
        =\left(D_\mu\big|H|^2\right)
        = 2 \mathcal{O}_H,\\[0.45\baselineskip]
 & \text{tr}\big[\big( H H^\dagger \big)P^2 \big( H H^\dagger \big)\big]
        = \text{tr}\big[ D_\mu\big(H H^\dagger \big)D^\mu \big( H H^\dagger \big)\big] 
        = \mathcal{O}_H + \mathcal{O}_T + 2 \mathcal{O}_R,\\[0.45\baselineskip]
 &\text{tr}\big[\big( \tilde{H}^\dagger (\tilde{H}^\dagger)^T\big)P^2 \big( \tilde{H} (\tilde{H})^T\big)\big]
        =\text{tr}\big[D_\mu\big( \tilde{H}^\dagger (\tilde{H}^\dagger)^T\big)D^\mu \big( \tilde{H} (\tilde{H})^T\big)\big] 
        = \mathcal{O}_H - \mathcal{O}_T + 2 \mathcal{O}_R ,\\
&H^\dag \overset\leftrightarrow{D_{\mu}} H\equiv H^\dag(D_\mu H)-(D_\mu H)^\dag H , \quad  H^\dag \tau^a \overset\leftrightarrow{D_{\mu}} H\equiv H^\dag\tau^a(D_\mu H)-(D_\mu H)^\dag \tau^a H , \label{app:essential2}
    \end{align}
where all the dimension six pure bosonic operators in Eqs.~\eqref{app:dim1}-\eqref{app:dim2} and Eqs.~\eqref{app:essential1}-\eqref{app:essential2} are listed in Tab.\,\ref{tbl:operators}.
\end{appendices}

\bibliographystyle{JHEP} 
\bibliography{main}

\end{document}